\documentclass[journal]{IEEEtran}
\usepackage{cite}
\usepackage{graphicx}
\usepackage{amsmath}
\usepackage{times}
\usepackage{latexsym}
\usepackage{graphicx}
\usepackage{bm}
\usepackage{amssymb}
\usepackage[center]{caption2}
\usepackage{stfloats}
\usepackage{cases}
\usepackage{array}
\usepackage{setspace}
\usepackage{fancyhdr}
\usepackage{citesort}
\usepackage{color}
\usepackage{subfigure}

\renewcommand{\QED}{\QEDopen}
\newtheorem{theorem}{Theorem}

\def\proof{\noindent\hspace{2em}{\itshape Proof: }}
\def\endproof{\hspace*{\fill}~$\square$\par\endtrivlist\unskip}

\begin{document}
\newpage
\title{Ergodic Capacity Comparison of Different Relay Precoding Schemes in Dual-Hop AF Systems with Co-Channel Interference}



\author{Guangxu Zhu,~\IEEEmembership{Student Member,~IEEE,} Caijun Zhong,~\IEEEmembership{Member,~IEEE,} Himal A. Suraweera,~\IEEEmembership{Member,~IEEE,}  Zhaoyang Zhang,~\IEEEmembership{Member,~IEEE,} Chau Yuen~\IEEEmembership{Member,~IEEE} and Rui Yin,~\IEEEmembership{Member,~IEEE}
\thanks{G. Zhu, C. Zhong, Z. Zhang, R. Yin are with the Institute of Information and Communication Engineering, Zhejiang University, China. (email: caijunzhong@zju.edu.cn).}
\thanks{H. A. Suraweera is with the Department of Electrical \& Electronic
Engineering, University of Peradeniya, Peradeniya 20400, Sri Lanka (email:
himal@ee.pdn.ac.lk).}
\thanks{Chau Yuen is with the Singapore University of Technology and Design, 20 Dover
Drive, Singapore 138682 (email: yuenchau@sutd.edu.sg)}}

\maketitle

\begin{abstract}
In this paper, we analyze the ergodic capacity of a dual-hop amplify-and-forward relaying system where the relay
is equipped with multiple antennas and subject to co-channel interference (CCI) and the additive white Gaussian noise. Specifically,
we consider three heuristic precoding schemes, where the relay first applies the 1) maximal-ratio combining (MRC) 2) zero-forcing (ZF) 3) minimum mean-squared error (MMSE) principle to combine the signal from the source, and then steers the transformed signal towards the destination with the maximum ratio transmission (MRT) technique. For the MRC/MRT and MMSE/MRT schemes, we present new tight analytical upper and lower bounds for the ergodic capacity, while for the ZF/MRT scheme, we derive a new exact analytical ergodic capacity expression. Moreover, we make a comparison among all the three schemes, and our results reveal that, in terms of the ergodic capacity performance, the MMSE/MRT scheme always has the best performance and the ZF/MRT scheme is slightly inferior, while the MRC/MRT scheme is always the worst one. Finally, the asymptotic behavior of ergodic capacity for the three proposed schemes are characterized in large $N$ scenario, where $N$ is the number of relay antennas. Our results reveal that, in the large $N$ regime, both the ZF/MRT and MMSE/MRT schemes have perfect interference cancelation capability, which is not possible with the MRC/MRT scheme.
\end{abstract}

%

\section{Introduction}\label{section:1}
Although decades of advancements in communication theory and practice have vastly empowered current cellular systems with improved performance, providing satisfactory throughput in the cell edge region is still a major challenge \cite{AKYILDIZ,GUO}. Towards this end, one effective solution that has received wide acceptance is the deployment of wireless relays \cite{CHO}. One or more relays are implemented in a network to assist the communication between the source and destination. Two popular relaying protocols that have been extensively studied in the literature are amplify-and-forward (AF) and decode-and-forward (DF) \cite{HASNA,TSIFTSIS}. An AF relay mimics the simple repeater functionality by amplifying the received signal, while the DF relay on the other hand decodes the source messages and forwards them to the destination.

To gain a fundamental understanding on the performance of relaying systems, a great deal of works have investigated the Shannon capacity in various practical relaying systems. For single antenna systems, the ergodic capacity of fixed-gain and variable gain relaying with an arbitrary number of relays in Rayleigh fading was studied in \cite{FARHADI}, and closed-form approximations and bounds of fixed-gain AF relaying systems in more general fading models were presented in later works, including Nakagami-$m$ fading \cite{COSTA} and $\mathcal{G}$-fading \cite{Zhong_TVT}. In \cite{FAN}, the authors derived an exact ergodic capacity expression for the variable-gain AF relaying system over Rayleigh fading channels. Several authors have also looked at the ergodic capacity of multi-antenna AF relaying systems. Using finite-dimensional random matrix theory, \cite{SHI} investigated the capacity of multiple-input multiple-output (MIMO) AF dual-hop systems with arbitrary finite antenna configurations, while in \cite{FIRAG}, an ergodic capacity analysis of MIMO AF channels with direct link between the source and destination was presented. It is worth pointing out that all these above works assume an interference free environment.

Due to the spectrum scarcity, future generations of commercial wireless systems are likely to adopt an aggressive frequency reuse policy in order to meet increasing demand for high quality wireless services. As such, relays deployed in 4G systems such as 3GPP LTE-Advanced, 802.16 j/m and IMT-Advanced can be subject to co-channel interference (CCI) from simultaneous transmissions on the same frequency channel \cite{GUO}. The presence of CCI can severely degrade the system performance as demonstrated in a rich body of publications on the performance of dual/multi-hop relay systems systems with CCI. For example, the detrimental effect of CCI on the outage probability of AF relay systems has been examined in various fading models and communication scenarios, including Rayleigh and Nakagami-$m$ fading \cite{ZHONG_TCOM1,FAWAZ}, single/multiple interferer with different cases of interference at the relay and/or the destination \cite{SALAMA,HIMAL_TVT}, relay selection \cite{KRIKIDIS} and multiple antenna systems
\cite{ZHONG_TCOM2,ZHU_TWC}.

On the other hand, so far, only few papers have investigated the capacity of AF relaying systems in the presence of CCI. For single antenna systems, a closed-form expression for the ergodic capacity of a dual-hop system equipped with a single fixed-gain relay subject to interference was derived in \cite{WAQAR}. The capacity of dual-hop and multi-hop AF relaying systems over Nakagami-$m$ fading with interference limited conditions was examined in \cite{TRIGUI1,TRIGUI2}. These studies have shed insights into how the performance of the system is affected by the dominant CCI factors, including the interference power and the fading severity. For multi-antenna cases, considering feedback delay and CCI, the ergodic capacity of a transmit beamforming/maximum ratio combining (MRC) AF dual-hop system equipped with a single antenna relay was studied in \cite{HUANG}. In \cite{ZHU_CONF}, assuming short-term/long-term relay power constraints, the ergodic capacity of a CCI impaired dual-hop system with zero-forcing (ZF)/maximal radio transmission (MRT) processing at the multi-antenna relay was investigated. This contemporary list of reference suggests that, while some progress has been made, significant efforts are required to gain a thorough understanding on the effect of multiple antennas with linear processing on the fundamental capacity limits of dual-hop systems with CCI.

Motivated by this, we consider a multiple antenna AF dual-hop system with interference at the relay. We adopt a system model where the relay is equipped with multiple antennas while the source and the destination have a single antenna each. This particular scenario is applicable in device-to-device (D2D) communication over cellular architecture, where due to the unavailability of a strong direct link, two low complexity device nodes select a sophisticated multi-antenna base-station to carry relayed traffic. The interference at the relay is a widely assumed assumption in the literature, and could also appear in practice where the source-relay link and the relay-destination link occupy different frequency bands, hence experience different interference patterns.


It is well known that, with multiple antennas, linear processing techniques attain desirable tradeoff between the implementation complexity and system performance and are very effective methods to combat the CCI. As such, in this paper, we investigate the impact of linear processing schemes on the ergodic capacity of dual-hop AF systems with CCI. Specifically, apart from the ZF/MRT scheme studied in \cite{ZHU_CONF}, we also consider another two popular linear processing techniques\cite{W.Zhang}, i.e., the maximum ratio combining (MRC)/maximal ratio transmission (MRT) scheme and the minimum mean square error (MMSE)/MRT scheme, and present a detailed study of all the three considered schemes. Our main contributions are summarized as follows:

\begin{itemize}
  \item For the MRC/MRT scheme and the MMSE/MRT scheme, we present analytical upper and lower bounds for the ergodic capacity of the system. These bounds remains sufficiently tight across the entire SNR range of interest, hence provide an efficient means for the evaluation of the ergodic capacity.
  \item For the ZF/MRT scheme, we present an exact analytical expression for the ergodic capacity of the system.
  \item {We also look into the asymptotic large $N$ regime, where the MMSE/MRT and the ZF/MRT achieve the same ergodic capacity which is identical to the system without CCI, and present an exact expression for the ergodic capacity.}
  \item {Our results suggest that, among three schemes considered, the MMSE/MRT scheme attains the highest ergodic capacity and the ZF/MRT scheme is slightly inferior, while the MRC/MRT scheme is the worst one.} In addition, increasing the number of relay antennas significantly enhance the ergodic capacity. Moreover, we examine numerically the impact of interference power distribution on the MMSE/MRT scheme, and it was demonstrated that the equal interference power scenario results in the lowest ergodic capacity.
 \end{itemize}

The rest of the paper is organized as follows: Section II introduces the system model. Section III presents the
exact or upper/lower bound analytical expressions for the ergodic capacity of the three linear processing schemes. Numerical results
and discussions are provided in Section IV. Finally, Section V concludes the paper and summarizes the main findings.

{\it Notation}: We use bold upper case letters to denote matrices, bold lower case letters to denote vectors and lower case
letters to denote scalers. {${\left\| {\bf{h}} \right\|_F}$ denotes the Frobenius norm, ${\tt E}\{x\}$ stands for the expectation of
random variable $x$, ${*}$ denotes the conjugate operator, while $T$ denotes the transpose operator and ${\dag}$ denotes the conjugate transpose operator. ${{\bf{I}}_M}$ is the identity matrix of size M. ${\sf diag}(\cdot)$ denotes the diagonal matrix. $n!$ denotes the factorial of integer $n$ and $\Gamma(x)$ is the gamma function. $\Gamma \left( {\alpha ,x} \right)$ is the upper incomplete gamma function \cite[Eq. (8.350.2)]{Tables}, $\psi(x)$ is the digamma function \cite[Eq. (8.360.1)]{Tables}, $\Psi \left( {a,b;z} \right)$ is the confluent hypergeometric function \cite[Eq. (9.210.2)]{Tables},  ${\Psi ^{\left( {1,0,0} \right)}}\left( {a,b;z} \right)$ denotes the derivative of $\Psi \left( {a,b;z} \right)$ with respect to $a$, and ${\Psi ^{\left( {0,1,0} \right)}}\left( {a,b;z} \right)$ denotes the derivative of $\Psi \left( {a,b;z} \right)$ with respect to $b$. Both the functions are available in popular softwares such as MATHEMATICA. $K_v(x)$ is the $v$-th order modified Bessel function of the second kind \cite[Eq. (8.407.1)]{Tables}.  ${\mathop{\rm G}\nolimits} \left( { \cdot \left|  \cdot  \right.} \right)$ is the Meijer's G function \cite[Eq. (9.301)]{Tables}  and ${\mathop{\rm G}\nolimits} _{1,[1:1],0,[1:1]}^{1,1,1,1,1}\left(  \cdot  \right)$ denotes the generalized Meijer's G-function of two variables \cite{R.P.Agrawal} which can be computed by the algorithm presented in \cite[Table II]{I.S.Ansari}. ${}_2{F_1}(a,b;c;z)$ is the Gauss hypergeometric function \cite[Eq. (9.100)]{Tables}. ${{\cal CN} (0,1)}$ denotes a scalar complex Gaussian distribution with zero mean and unit variance.

\section{System Model}\label{section:2}
Fig. \ref{fig:fig1} shows the dual-hop AF relaying system considered in this paper. Because of size and complexity constraints, the source and the destination is only equipped with one antenna, while the more sophisticated relay, e.g., a base-station has multiple antennas. An interference scenario in which the relay is subjected to $M$ independently but not necessarily identically distributed co-channel interferers and additive white Gaussian noise (AWGN), while the destination is corrupted by AWGN only is assumed.\footnote{{Please note, the analysis of the MRC/MRT and MMSE/MRT schemes presented in the ensuing section can be extend to the general scenario where both the relay and destination are subject to CCI. However, since the main purpose of the current work is to study the effect of multiple antennas on combating the CCI, considering the CCI at the destination would only complicate the analysis, yet providing no additional insight. Hence, we limit ourself to the scenario where only the relay node is subject to CCI.}} In this dual-hop system, the direct link is very weak and ignored due to high shadowing and path loss between the source and the destination.
\begin{figure}[ht]
\centering
\includegraphics[scale=0.8]{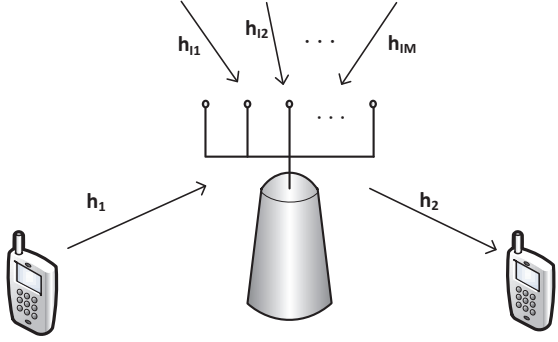}
\caption{A schematic diagram of the system model.}\label{fig:fig1}
\end{figure}

In the considered dual-hop system, due to the half-duplex constraint, total communication between the source and the destination takes place in two
time slots. In the first time slot, the source sends its signal to the relay and the received signal at the relay can be expressed as
\begin{align}\label{SM:1}
{{{\bf y}}_{{r}}} = {{{\bf h}}_{{1}}}x + \sum\limits_{i = 1}^M {{{{\bf h}}_{{Ii}}}{s_{Ii}} + {{{\bf n}}_{{1}}}} ,
\end{align}
where the channel gain for the source-relay link denoted by ${{{\bf h}}_{{1}}}$ is an ${N\times 1}$ vector, and its entries follow identically and independently distributed (i.i.d.) ${{\cal CN} (0,1)}$, the channel gain for the \emph{i}-th interference-relay link denoted by ${{{\bf h}}_{{Ii}}}$ is an ${N\times 1}$ vector, and its entries follow  i.i.d. ${{\cal CN} (0,1)}$, $x$ is the source symbol satisfying ${\tt E}\left\{ {x{x^* }} \right\} = P$. ${s_{Ii}}$ is the \emph{i}-th interference symbol with ${\tt E}\left\{ {{s_{Ii}}s_{Ii}^*} \right\} = {P_{Ii}}$, ${{{\bf n}}_{{1}}}$ is an ${N\times 1}$ vector and denotes the AWGN at the relay node with ${\tt E}\{\mathbf{n}_1\mathbf{n}_1^{\dag}\} = N_0{\mathbf{I}}$.

A linear procoder is applied to the received signal in (1) and transmitted to the destination in the second time slot. Therefore the scalar received
signal at the destination can be written as
\begin{align}\label{SM:2}
y_d = {\bf{h}}^{\dag}_{{2}}{\bf{W}}{{\bf{y}}_{{r}}} +n_2,
\end{align}
where the channel gain for the relay-destination link denoted by ${{\bf{h}}_{{2}}}$ is a ${N\times 1}$ vector, and its entries follow i.i.d. ${{\cal CN} (0,1)}$, ${n_2}$ is the AWGN at destination with ${\tt E}\{n_2^{*}n_2\} = N_0$, $\mathbf{W}$ is the transformation matrix at relay node with ${\tt E}\{|\mathbf{W}\mathbf{y}_r|^2\} = P_r$.

Invoking (\ref{SM:1}) and (\ref{SM:2}), the end-to-end signal-to-interference-and-noise ratio (SINR) of the system can be computed as
\begin{align}\label{SINR:1}
\gamma  = \frac{{{\left| {{\bf{h}}^{\dag}_{2}{\bf{W}}{{\bf{h}}_{1}}} \right|}^2}P}{{\sum\limits_{i = 1}^M {{\left| {{{\bf h}}^{\dag}_{2}{\bf{W}}{{{\bf h}}_{Ii}}} \right|}^2}{P_{Ii}} + {\| {{{\bf h}}^{\dag}_{{2}}{\bf{W}}} \|_F}^2}{N_0} + {N_0}}.
\end{align}

In general, due to the non-convex nature of the problem, the optimal relay transformation matrix ${\bf W}$ maximizing the end-to-end SINR $\gamma$ does not seem to be analytically tractable. Hence, in this paper, a two-stage relay processing strategy is considered, i.e., the relay first utilizes linear processing methods to suppress the CCI, and then forwards the transformed signal to the destination using the MRT scheme. As such, the matrix ${\bf W}$ is a rank-1 matrix, which can be expressed as ${\bf{W}} = \omega \frac{{{\bf{h}}_2}}{{\| {{{\bf{h}}_2}} \|_F}}{{\bf{w}}_1}$, where $\omega$ is the power constraint factor, $\frac{{{\bf{h}}_2}}{{\| {{{\bf{h}}_2}} \|_F}}$ is the MRT precoder and ${{\bf{w}}_1}$ is a ${1\times N}$ linear combining vector, which depends on the linear combining scheme employed by the relay. Specifically, here we consider three different linear combining schemes, namely, the MRC, the ZF and the MMSE schemes as detailed below. For notational convenience, we define ${\rho _1} = \frac{{P}}{{{N_0}}}$, ${\rho _2} = \frac{{{P_r}}}{{{N_0}}}$ and ${\rho _{Ii}} = \frac{{{P_{Ii}}}}{{{N_0}}}$, $i=1,\dots,M$.
\subsection{MRC Scheme}
The MRC scheme adds together all the signals received from each antenna to achieve a higher SNR, mathematically, the MRC combiner is given by ${{\bf{w}}_{1}}=\frac{{{\bf{h}}_1^{\bf{\dag }}}}{{\| {{{\bf{h}}_1}} \|_F}}$.
To meet the transmit power constraint at the relay, the constant factor $\omega^2$ can be computed as
{\begin{align}
{\omega ^2} = \frac{{{\rho_2}}}{{{{\bf h}}_{{1}}^{\bf{\dag }}{{{\bf h}}_{{1}}}\rho_1 + \frac{{\sum\limits_{i = 1}^M {{{\left| {{{\bf h}}_{{1}}^{{\dag }}{{{\bf h}}_{{{Ii}}}}} \right|}^2}{\rho_{Ii}}} }}{{{{\left\| {{{{\bf h}}_{{1}}}} \right\|}_F^2}}} + 1}},
\end{align}}
thus, the corresponding end-to-end SINR for the MRC/MRT scheme $\gamma_{\sf MRC}$ can be expressed as
\begin{align}\label{C:2}
\gamma_{\sf MRC} = \frac{{{\gamma _1^{\sf MRC}}{\gamma _2^{\sf MRC}}}}{{{\gamma _1^{\sf MRC}} + {\gamma _2^{\sf MRC}} + 1}},
\end{align}
where ${\gamma _1^{\sf MRC}} = \frac{{{{\| {{{\bf{h}}_{1}}} \|}^2_F}{\rho _1}}}{{{U_1} + 1}}$, ${U_1} = \sum\limits_{i = 1}^M {\frac{{{{\left| {{\bf{h}}_{\bf{1}}^\dag {{\bf{h}}_{{\bf{Ii}}}}} \right|}^2}}}{{{{\| {{{\bf{h}}_{\bf{1}}}} \|}^2_F}}}{\rho _{Ii}}} $, ${\gamma_2^{\sf MRC}} = {\| {{{\bf{h}}_2}} \|^2_F\rho_2}$.

It is well known that with independent fading at each antenna element in the presence of spatially AWGN, the MRC scheme is optimal in terms of maximizing the end-to-end SNR. However, in the presence of interference, MRC is in general suboptimal, as it treats the interference as noise. Hence, it is of great interest to look at more sophisticated linear combining schemes with superior interference suppression capability, i.e., the ZF or the MMSE scheme.
\subsection{ZF Scheme}
The ZF scheme intends to completely eliminate the CCI. To ensure this is possible, the number of the antennas equipped at the relay should be greater than the number of interferers. Hence, for the ZF/MRT scheme, it is assumed that $N > M$. According to \cite[Proposition 1]{ZHU_TWC}, the optimal ZF combining vector ${\bf{w}}_1$ is given by
\begin{align}
{\bf{w}}_1 = \frac{{{\bf{h}}_1^\dag {\bf{P}}}}{{\sqrt {{\bf{h}}_1^\dag {\bf{P}}{{\bf{h}}_1}} }},
\end{align}
where ${\bf{P}} = {{\bf{I}}_N} - {{\bf{H}}_I}{\left( {{\bf{H}}_I^\dag {{\bf{H}}_I}} \right)^{ - 1}}{\bf{H}}_I^\dag $ and ${{\bf{H}}_I} = \left[ {{{\bf{h}}_{I1}},{{\bf{h}}_{I2}} \cdots {{\bf{h}}_{IM}}} \right]$. Then, the power constraint factor can be calculated as
{\begin{align}
{\omega ^2} = \frac{{{\rho_2}}}{{\left| {{\bf{w}}_1{{\bf{h}}_1}} \right|^2{\rho_1} + 1}}.
\end{align}}

Therefore, the corresponding end-to-end SINR of the ZF/MRT scheme can be written as
\begin{align}\label{C:7}
\gamma_{\sf ZF} = \frac{{{\gamma _1^{\sf ZF}}{\gamma _2^{\sf ZF}}}}{{{\gamma _1^{\sf ZF}} + {\gamma _2^{\sf ZF}} + 1}},
\end{align}
where ${\gamma _1^{\sf ZF}} = {\left| {{\bf{h}}_1^\dag {\bf{P}}{{\bf{h}}_1}} \right|}{\rho _1}$, ${\gamma _2^{\sf ZF}} = {{{\| {{{\bf{h}}_{2}}} \|}^2_F}}{\rho _2}$.

\subsection{MMSE Scheme}
The ZF scheme completely eliminates the CCI at the relay, which however causes an elevated noise level. In contrast, the MMSE scheme does not fully eliminate the CCI, instead, it provides the optimum trade-off between interference suppression and noise enhancement. To make the analysis tractable, we assume that ${\rho _{Ii}} \equiv {\rho _I},\;\forall i = 1,2 \ldots M $, thus, we also have ${P_{Ii}} \equiv {P_I},\;\forall i = 1,2 \ldots M$. According to \cite{H.Gao}, the MMSE combiner should be set as ${{\bf{w}}_1} = {\bf{h}}_1^\dag {\left( {{{\bf{h}}_1}{\bf{h}}_1^\dag  + {{\bf{H}}_I}{\bf{H}}_I^\dag  + \frac{{{N_0}}}{{{P_I}}}{\bf{I}}_N} \right)^{ - 1}}$. {It is important to note that there exists some practical scenarios where the equal interference power assumption adopted to simplify the analytical derivation becomes realistic.} For example, it applies when the interference sources are clustered together \cite{D.Costacluster,K.Gulati} or when the interference originates from a multiple antenna source implementing an uniform power allocation policy. In addition, we will later illustrate numerically in Section \ref{section:4} that our analytical results in Section \ref{section:3} provide very accurate approximations to the ergodic capacity for scenarios with distinct interference power.


Also, in order to meet the power constraint at the relay, we have
{\begin{align}
{\omega ^2} = \frac{{{\rho_2}}}{{{{\left| {{\bf{w}}_1 {{\bf{h}}_1}} \right|}^2}\rho_1 + \sum\limits_{i = 1}^M {{{\left| {{\bf{w}}_1 {{\bf{h}}_{Ii}}} \right|}^2}} {\rho_I} + {{\left\| {{\bf{w}}_1 } \right\|}_F^2}}}.
\end{align}}
Therefore, the corresponding end-to-end SINR for the MMSE/MRT scheme can be expressed as
\begin{align}\label{MMSE:2}
\gamma_{\sf MMSE} = \frac{{{\gamma _1^{\sf MMSE}}{\gamma _2^{\sf MMSE}}}}{{{\gamma _1^{\sf MMSE}} + {\gamma _2^{\sf MMSE}} + 1}},
\end{align}
where ${\gamma _1^{\sf MMSE}} = \frac{{{P_s}}}{{{P_I}}}{\bf{h}}_1^\dag {{\bf{R}}^{ - 1}}{{\bf{h}}_1}$, ${\bf{R}} = {{\bf{H}}_I}{\bf{H}}_I^\dag  + \frac{{{N_0}}}{{{P_I}}}{\bf{I}}_N$ and ${\gamma_2^{\sf MMSE}} = {\| {{{\bf{h}}_2}} \|^2_F\rho_2}$.

{\it Remark}: {We would like to point out that the channel state information (CSI) requirement is different for the considered three schemes. Specifically, the MRC/MRT scheme only requires the knowledge of ${\bf h}_1$ and ${\bf h}_2$, the ZF/MRT scheme requires the knowledge of ${\bf h}_1$, ${\bf h}_2$, and ${\bf H}_I$, while the MMSE/MRT scheme has the highest CSI requirement, since the noise variance $N_0$ at the relay is also needed besides the knowledge of ${\bf h}_1$, ${\bf h}_2$, and ${\bf H}_I$. Please note, the CSI of CCI can be obtained by utilizing the methods given in the literature \cite{D.Katselis,Y.Ohwatari,R.Narasimhan}. In general, if more CSI is available at the transmitter, more sophisticated transmission schemes could be designed to improve the system performance. However, more CSI also implicitly implies a higher system overhead. Therefore, when designing practical wireless systems, it is important to take this tradeoff into consideration.}

\section{Ergodic Capacity Analysis}\label{section:3}
In this section, we present a rigorous investigation on the ergodic capacity of the MRC/MRT, ZF/MRT and MMSE/MRT schemes introduced in Section \ref{section:2}.  Mathematically, the ergodic capacity is defined as the expected value of the instantaneous mutual information, and it can be given by\footnote{It is assumed that the source
and all the interferers use the Gaussian signaling. Without CSI at the source, adopting the Gaussian
signaling is a reasonable choice, and such assumption has been widely adopted in the literature, see
for instance \cite{M. Chiani}.}
\begin{align}\label{C:1}
C = \frac{1}{2}{{\mathop{\rm E}\nolimits}}\left[ {{{\log }_2}\left( {1 + {\gamma}} \right)} \right],
\end{align}
where $\gamma$ is the end-to-end SINR of the system and the factor $1/2$ accounts for the fact that the entire communication occupies two time slot.

\subsection{MRC/MRT Scheme}
The ergodic capacity of the MRC/MRT scheme is given by
\begin{align}\label{C:3}
C_{\sf MRC} = \frac{1}{2}{{\mathop{\rm E}\nolimits} }\left[ {{{\log }_2}\left( {1 + {\gamma_{\sf MRC}}} \right)} \right],
\end{align}
where $\gamma_{\sf MRC}$ is given in (\ref{C:2}). Unfortunately, exact evaluation of the ergodic capacity in (\ref{C:3}) is in general impossible, since the cumulative distribution function (c.d.f) of (\ref{C:2}) can not be given in closed-form. Motivated by this, we hereafter seek to deduce upper and lower
bounds on $C_{\sf MRC}$.

Substituting (\ref{C:2}) into (\ref{C:3}), the ergodic capacity of the MRC/MRT scheme can be expressed as
\begin{align}\label{C:4}
C_{\sf MRC}  &= \frac{1}{2}{\mathop{\rm E}\nolimits} \left[ {{{\log }_2}\left( {\frac{{\left( {1 + {\gamma _1^{\sf MRC}}} \right)\left( {1 + {\gamma _2^{\sf MRC}}} \right)}}{{1 + {\gamma _1^{\sf MRC}} + {\gamma _2^{\sf MRC}}}}} \right)} \right]\notag\\
 &= C_{\gamma _1^{\sf MRC}}+C_{\gamma _2^{\sf MRC}}-C_{\gamma _T^{\sf MRC}},
\end{align}
where $C_{\gamma _i^{\sf MRC}} = \frac{1}{2}{\mathop{\rm E}\nolimits} \left[ {{{\log }_2}\left( {1 + {\gamma _i^{\sf MRC}}} \right)} \right]$, for $i \in \{ 1,2\} $, and $C_{\gamma _T^{\sf MRC}} = \frac{1}{2}{\mathop{\rm E}\nolimits} \left[ {{{\log }_2}\left( {1 + {\gamma _1^{\sf MRC}}+ {\gamma _2^{\sf MRC}}} \right)} \right]$. A direct evaluation of $C_{\gamma _T^{\sf MRC}}$ does not seem to be possible due to the difficulty in obtaining closed-form expression for the c.d.f. of ${\gamma _1^{\sf MRC}}+ {\gamma _2^{\sf MRC}}$. Hence, we seek a tight bound in the following.
Noticing that $f\left( {x,y} \right) = {\log _2}\left( {1 + {e^x} + {e^y}} \right)$ is a convex function with respect to $x$ and $y$, we have
\begin{align}\label{C:5}
{C_{{\gamma _T^{\sf MRC}}}} \ge \frac{1}{2}{\log _2}\left( {1 + {e^{{\mathop{\rm E}\nolimits} \left( {\ln {\gamma _1^{\sf MRC}}} \right)}} + {e^{{\mathop{\rm E}\nolimits} \left( {\ln {\gamma _2^{\sf MRC}}} \right)}}} \right).
\end{align}
With the help of (\ref{C:5}), we establish the ergodic capacity upper bound in the following theorem:
\begin{theorem}\label{theorem:1}
The ergodic capacity of the MRC/MRT scheme is upper bounded by
\begin{multline}
{C_{\sf MRC}^{\sf up}} = \frac{{{\rho _1}}}{{2\ln 2}}\sum\limits_{k = 0}^{N - 1} {\frac{1}{{k!}}\sum\limits_{l = 0}^k {k\choose l} } \sum\limits_{i = 1}^{\rho ({\bf{D}})} {\sum\limits_{j = 1}^{{\tau _i}({\bf{D}})} {{\chi _{i,j}}({\bf{D}})\frac{{\rho _{I\left\langle i \right\rangle }^l}}{{\Gamma \left( j \right)}}}}\\
{\mathop{\rm G}\nolimits} _{1,[1:1],0,[1:1]}^{1,1,1,1,1}\left(^{{\rho _1}}_{{\rho _{I\left\langle i \right\rangle }}}
\middle| \substack{
{k + 1}\\
{0;1 - j - l}\\
 - \\
{0;0}} \right)
 + \frac{{e^{\frac{1}{{{\rho _2}}}}}}{{2\ln 2}}\sum\limits_{k = 0}^{N - 1} {{{\frac{1}{{{\rho _2^k}}}}}\Gamma \left( { - k,\frac{1}{{{\rho _2}}}} \right)}  \\- \frac{1}{2}{\log _2}\left( {1 + {\rho _2}\exp \left( {\psi \left( N \right)} \right) + \exp \left( {{\cal A}_1} \right)} \right),
\end{multline}
where ${\cal A}_1$ is given by (\ref{C:5.1}) shown on the top of the next page,
\begin{figure*}
\begin{multline}\label{C:5.1}
{\cal A}_1 = \sum\limits_{i = 1}^{\rho ({\bf{D}})} {\sum\limits_{j = 1}^{{\tau _i}({\bf{D}})} {{\chi _{i,j}}({\bf{D}})\left[ {\left( {\ln \frac{{{\rho _1}}}{{{\rho _{I\left\langle i \right\rangle }}}} + \psi \left( 1 \right)} \right)\Psi \left( {0,1 - j;\frac{1}{{{\rho _{I\left\langle i \right\rangle }}}}} \right)} \right.} }
 + \left. {{\Psi ^{\left( {1,0,0} \right)}}\left( {0,1 - j;\frac{1}{{{\rho _{I\left\langle i \right\rangle }}}}} \right) + {\Psi ^{\left( {0,1,0} \right)}}\left( {0,1 - j;\frac{1}{{{\rho _{I\left\langle i \right\rangle }}}}} \right)} \right]\\
 + \sum\limits_{k = 1}^{N - 1} {\frac{1}{k}\sum\limits_{l = 0}^k {k\choose l} } \sum\limits_{i = 1}^{\rho ({\bf{D}})} {\sum\limits_{j = 1}^{{\tau _i}({\bf{D}})} {{\chi _{i,j}}({\bf{D}})\frac{{\Gamma \left( {j + l} \right)}}{{\Gamma \left( j \right)}}\rho _{I\left\langle i \right\rangle }^{l - k}\Psi \left( {k,k - j - l + 1;\frac{1}{{{\rho _{I\left\langle i \right\rangle }}}}} \right)} },
\end{multline}
\hrule
\end{figure*}
 $\mathbf{D} = {\sf diag}({\rho _{I1}},{\rho _{I2}}, \cdots ,{\rho _{IM}})$, $\rho (\mathbf{D})$ is the number of distinct diagonal elements of $\mathbf{D}$, ${\rho _{I\left\langle 1 \right\rangle }} > {\rho _{I\left\langle 2 \right\rangle }} >  \cdots  > {\rho _{I\left\langle {\rho (\mathbf{D})} \right\rangle }}$ are the distinct diagonal elements in decreasing order, ${\tau _i}(\mathbf{D})$ is the multiplicity of ${\rho _{I\left\langle i \right\rangle }}$ and ${\chi _{i,j}}(\mathbf{D})$ is the ${\left( {i,j} \right)-\mbox{th}}$ characteristic coefficient of $\mathbf{D}$.

\proof See Appendix \ref{appendix:theorem:1}. \endproof
\end{theorem}

Now, let us consider the derivation for the lower bound. Applying the Jensen's inequality on $C_{\gamma _T^{\sf MRC}}$, we have
\begin{align}\label{C:6}
{C_{{\gamma _T^{\sf MRC}}}} \le \frac{1}{2}{\log _2}\left( {1 + {\mathop{\rm E}\nolimits} \left( {{\gamma _1^{\sf MRC}}} \right) + {\mathop{\rm E}\nolimits} \left( {{\gamma _2^{\sf MRC}}} \right)} \right).
\end{align}
According to (\ref{C:6}), we have the following key result:
\begin{theorem}\label{theorem:2}
The ergodic capacity of the MRC/MRT scheme is lower bounded by
\begin{multline}
{C_{\sf MRC}^{\sf low}} = \frac{{{\rho _1}}}{{2\ln 2}}\sum\limits_{k = 0}^{N - 1} {\frac{1}{{k!}}\sum\limits_{l = 0}^k {k\choose l} } \sum\limits_{i = 1}^{\rho ({\bf{D}})} {\sum\limits_{j = 1}^{{\tau _i}({\bf{D}})} {{\chi _{i,j}}({\bf{D}})}}\\
\times\frac{{\rho _{I\left\langle i \right\rangle }^l}}{{\Gamma \left( j \right)}}{\mathop{\rm G}\nolimits} _{1,[1:1],0,[1:1]}^{1,1,1,1,1}\left(^{{\rho _1}}_{{\rho _{I\left\langle i \right\rangle }}}
\middle| \substack{
{k + 1}\\
{0;1 - j - l}\\
 - \\
{0;0}} \right) +\\
 \frac{{e^{\frac{1}{{{\rho _2}}}}}}{{2\ln 2}}\sum\limits_{k = 0}^{N - 1} {{{\frac{1}{{{\rho _2^k}}}}}\Gamma \left( { - k,\frac{1}{{{\rho _2}}}} \right)}  - \frac{1}{2}{\log _2}\left( {1 + N{\rho _2} + {\cal A}_2} \right),
\end{multline}
with
\begin{multline}
{\cal A}_2 = {\rho _1}\sum\limits_{k = 0}^{N - 1} {\sum\limits_{l = 0}^k {\left( {\begin{array}{*{20}{c}}
k\\
l
\end{array}} \right)} } \sum\limits_{i = 1}^{\rho ({\bf{D}})} {\sum\limits_{j = 1}^{{\tau _i}({\bf{D}})} {{\chi _{i,j}}({\bf{D}})}}\frac{{\Gamma \left( {j + l} \right)}}{{\Gamma \left( j \right)}}\\
\times\rho _{I\left\langle i \right\rangle }^{l - k - 1}\Psi \left( {k + 1,k - j - l + 2;\frac{1}{{{\rho _{I\left\langle i \right\rangle }}}}} \right).
\end{multline}

\proof See Appendix \ref{appendix:theorem:2}. \endproof
\end{theorem}


\subsection{ZF/MRT Scheme}
Starting from (\ref{C:7}), the ergodic capacity is given in the following theorem:
\begin{theorem}\label{theorem:3}
The ergodic capacity of the ZF/MRT scheme can be expressed as (\ref{C:8}) shown on the top of the next page.
\begin{figure*}
\begin{multline}\label{C:8}
C_{\sf ZF} = \frac{1}{{2\ln 2}}\left( {\sum\limits_{k = 0}^{N - M - 1} {{\frac{{e^{\frac{1}{{{\rho _1}}}}}}{{{\rho _1^k}}}}}\Gamma \left( { - k,\frac{1}{{{\rho _1}}}} \right)} \right. + \sum\limits_{j = 0}^{N - 1}  {\frac{{e^{\frac{1}{{{\rho _2}}}}}}{{{\rho _2^j}}}}\Gamma \left( { - j,\frac{1}{{{\rho _2}}}} \right)- \sum\limits_{k = 0}^{N - M - 1} {\Psi \left( {1,1 - k;\frac{1}{{{\rho _1}}}} \right)}\\
- \sum\limits_{j = 0}^{N - 1} {\Psi \left( {1,1 - j;\frac{1}{{{\rho _2}}}} \right)}
   + \left. { {\rho _1}{\rho _2}\sum\limits_{k = 0}^{N - M - 1} {\sum\limits_{j = 0}^{N - 1} {{\mathop{\rm G}\nolimits} _{1,[1:1],0,[1:1]}^{1,1,1,1,1}\left(^{{\rho _1}}_{{\rho _2}}
\middle| \substack{
2\\
{ - k; - j}\\
 - \\
{0;0}} \right)} } } \right).
\end{multline}
\hrule
\end{figure*}

\proof See Appendix \ref{appendix:theorem:3}. \endproof
\end{theorem}

Theorem \ref{theorem:3} presents the exact analytical ergodic capacity expression of the ZF/MRT scheme, which is quite general and valid for the system with arbitrary number of antennas and interferers. Such an expression can be efficiently evaluated numerically using software such as
MATLAB or MATHEMATICA, which provides notable computational advantage over the Monte Carlo simulation method.

\subsection{MMSE/MRT Scheme}
Similar to the case in the MRC/MRT scheme, the exact ergodic capacity of the MMSE/MRT scheme $C_{\sf MMSE}$ is in general intractable. Hence, we hereafter try to deduce upper and lower bounds for $C_{\sf MMSE}$. It is easy to note that, the ergodic capacity of the MMSE/MRT scheme can be expressed as
\begin{align}\label{MMSE:3}
C_{\sf MMSE} &= \frac{1}{2}{\mathop{\rm E}\nolimits} \left[ {{{\log }_2}\left( {1 + \frac{{{\gamma _1^{\sf MMSE}}{\gamma _2^{\sf MRC}}}}{{{\gamma _1^{\sf MMSE}} + {\gamma _2^{\sf MMSE}} + 1}}} \right)} \right]\notag\\
 &= C_{\gamma _1^{\sf MMSE}}+C_{\gamma _2^{\sf MMSE}}-C_{\gamma _T^{\sf MMSE}},
\end{align}
where $C_{\gamma _i^{\sf MMSE}} = \frac{1}{2}{\mathop{\rm E}\nolimits} \left[ {{{\log }_2}\left( {1 + {\gamma _i^{\sf MMSE}}} \right)} \right]$, for $k \in \{ 1,2\} $, $C_{\gamma _T^{\sf MMSE}} = \frac{1}{2}{\mathop{\rm E}\nolimits} \left[ {{{\log }_2}\left( {1 + {\gamma _1^{\sf MMSE}}+ {\gamma _2^{\sf MMSE}}} \right)} \right]$.

Utilizing the same methods as in the case of the MRC/MRT scheme, we establish the upper and lower bounds as (\ref{MMSE:4}) shown on the top of the next page,
\begin{figure*}
\begin{align}\label{MMSE:4}
\frac{1}{2}{\log _2}\left( {1 + {e^{{\mathop{\rm E}\nolimits} \left( {\ln {\gamma _1^{\sf MMSE}}} \right)}} + {e^{{\mathop{\rm E}\nolimits} \left( {\ln {\gamma _2^{\sf MMSE}}} \right)}}} \right) \le {C_{{\gamma _T^{\sf MMSE}}}} \le \frac{1}{2}{\log _2}\left( {1 + {\mathop{\rm E}\nolimits} \left( {{\gamma _1^{\sf MMSE}}} \right) + {\mathop{\rm E}\nolimits} \left( {{\gamma _2^{\sf MMSE}}} \right)} \right).
\end{align}
\hrule
\end{figure*}
and we have the following result:
\begin{theorem}\label{theorem:4}
The ergodic capacity of the MMSE/MRT scheme is upper bounded by (\ref{MMSE:4.1}) shown on the top of the next page,
\begin{figure*}
\begin{multline}\label{MMSE:4.1}
{C_{\sf MMSE}^{\sf up}} = \frac{1}{{2\ln 2}}{e^{\frac{1}{{{\rho _1}}}}}\sum\limits_{k = 0}^{N - 1} {{{\left( {\frac{1}{{{\rho _1}}}} \right)}^k}\Gamma \left( { - k,\frac{1}{{{\rho _1}}}} \right)}  + \frac{1}{{2\ln 2}}{e^{\frac{1}{{{\rho _2}}}}}\sum\limits_{k = 0}^{N - 1} {{{\left( {\frac{1}{{{\rho _2}}}} \right)}^k}\Gamma \left( { - k,\frac{1}{{{\rho _2}}}} \right)} \\
 - \frac{{{\rho _1}}}{{2\ln 2}}\sum\limits_{m = m_1}^N {\frac{{\rho _I^{N - m + 2}}}{{\Gamma \left( m \right)\Gamma \left( {N - m + 1} \right)\Gamma \left( {m - N + M} \right)}}} {{\mathop{\rm G}\nolimits} _{1,[1:2],0,[1:2]}^{1,1,2,1,1}\left(^{{\rho _1}}_{{\rho _I}}\middle| \substack{
{N + 2}\\
{0;\left( { - M - 1,m - N - 1} \right)}\\
 - \\
{0;\left( { - 1,m - N - 2} \right)}} \right)}\\
 - \frac{1}{2}{\log _2}\left( {1 + {\rho _2}e^{\psi \left( N \right)} + {\rho _1}\exp \left( {\psi \left( N \right)-{\sum\limits_{m = {m_1}}^N {\frac{{\rho _I^{N - m + 2}} {\mathop{\rm G}\nolimits} _{3,2}^{1,3}\left( {\rho _I}\middle| \substack{
{ - N, - M - 1,m - N - 1}\\
{ - 1,m - N - 2}} \right)}{{\Gamma \left( m \right)\Gamma \left( {N - m + 1} \right)\Gamma \left( {m - N + M} \right)}}}}} \right)} \right),
\end{multline}
\hrule
\end{figure*}
where $m_1=\max \left( {0,N - M} \right) + 1$.

\proof See Appendix \ref{appendix:theorem:4}. \endproof
\end{theorem}

Now, we turn our attention to the ergodic capacity lower bound, and we have the following result.

\begin{theorem}\label{theorem:5}
The ergodic capacity of the MMSE/MRT scheme is lower bounded by (\ref{MMSE:4.2}) shown on the top of the next page.
\begin{figure*}
\begin{multline}\label{MMSE:4.2}
{C_{\sf MMSE}^{\sf low}} = \frac{1}{{2\ln 2}}{e^{\frac{1}{{{\rho _1}}}}}\sum\limits_{k = 0}^{N - 1} {{{\left( {\frac{1}{{{\rho _1}}}} \right)}^k}\Gamma \left( { - k,\frac{1}{{{\rho _1}}}} \right)}  + \frac{1}{{2\ln 2}}{e^{\frac{1}{{{\rho _2}}}}}\sum\limits_{k = 0}^{N - 1} {{{\left( {\frac{1}{{{\rho _2}}}} \right)}^k}\Gamma \left( { - k,\frac{1}{{{\rho _2}}}} \right)} \\
 - \frac{{{\rho _1}}}{{2\ln 2}}\sum\limits_{m = {m_1}}^N {\frac{{\rho _I^{N - m + 2}}}{{\Gamma \left( m \right)\Gamma \left( {N - m + 1} \right)\Gamma \left( {m - N + M} \right)}}} {{\mathop{\rm G}\nolimits} _{1,[1:2],0,[1:2]}^{1,1,2,1,1}\left(^{{\rho _1}}_{{\rho _I}}\middle| \substack{
{N + 2}\\
{0;\left( { - M - 1,m - N - 1} \right)}\\
 - \\
{0;\left( { - 1,m - N - 2} \right)}} \right)}\\
 - \frac{1}{2}{\log _2}\left( {1 + N{\rho _1} + N{\rho _2} - {\rho _1}\sum\limits_{m = {m_1}}^N {\frac{{\rho _I^{N - m + 2}}{{\mathop{\rm G}\nolimits} _{3,2}^{1,3}\left( {\rho _I}\middle| \substack{{ - N - 1, - M - 1,m - N - 1}\\
{ - 1,m - N - 2}} \right)}}{{\Gamma \left( m \right)\Gamma \left( {N - m + 1} \right)\Gamma \left( {m - N + M} \right)}}} } \right).
\end{multline}
\hrule
\end{figure*}

\proof See Appendix \ref{appendix:theorem:5}. \endproof
\end{theorem}



\subsection{Large N Analysis}
In this subsection, we look into the large $N$ regime with fixed $M$, and examine the asymptotic behavior of the proposed schemes. With the help of the law of large numbers, \cite{ZHU_TWC} has proven that, in the large $N$ regime, the end-to-end SINRs of both the ZF/MRT and MMSE/MRT schemes can be finally simplified to the exact end-to-end SNR of the same dual-hop AF relaying system but without CCI at the relay. It is given by
\begin{align}\label{LN:1}
\gamma^{\infty} = \frac{{{\gamma_1}{\gamma_2}}}{{{\gamma_1} + {\gamma_2} + 1}},
\end{align}
where $\gamma_1={\rho_1}{{{\| {{{\bf{h}}_{1}}} \|}^2_F}}$, and $\gamma_2={\rho_2}{{{\| {{{\bf{h}}_{2}}} \|}^2_F}}$. Please note, the large $N$ SINR approximation in (\ref{LN:1}) does not hold for the MRC/MRT scheme.  This is because that, for the MRC/MRT scheme, the effect of CCI persists regardless of the value of $N$.

Based on this key observation we have the following result.
\begin{theorem}\label{theorem:6}
When $N \to \infty $, the ergodic capacity of the ZF/MRT and MMSE/MRT schemes can be approximated as (\ref{LN:2}) shown on the bottom of the next page.
\begin{figure*}
\begin{multline}\label{LN:2}
C_{\sf LN} = \frac{1}{{2\ln 2}}\left( {\sum\limits_{k = 0}^{N  - 1} {{\left( {\frac{1}{{{\rho _1}}}} \right)}^k}{e^{\frac{1}{{{\rho _1}}}}}\Gamma \left( { - k,\frac{1}{{{\rho _1}}}} \right)} \right. + \sum\limits_{j = 0}^{N - 1}  {\left( {\frac{1}{{{\rho _2}}}} \right)^j}{e^{\frac{1}{{{\rho _2}}}}}\Gamma \left( { - j,\frac{1}{{{\rho _2}}}} \right)
 - \sum\limits_{k = 0}^{N  - 1} {\Psi \left( {1,1 - k;\frac{1}{{{\rho _1}}}} \right)} \\ - \sum\limits_{j = 0}^{N - 1} {\Psi \left( {1,1 - j;\frac{1}{{{\rho _2}}}} \right)}
   + \left. { {\rho _1}{\rho _2}\sum\limits_{k = 0}^{N - 1} {\sum\limits_{j = 0}^{N - 1} {{\mathop{\rm G}\nolimits} _{1,[1:1],0,[1:1]}^{1,1,1,1,1}\left(^{{\rho _1}}_{{\rho _2}}
\middle| \substack{2\\
{ - k; - j}\\
 - \\
{0;0}} \right)} } } \right).
\end{multline}
\hrule
\end{figure*}

\proof
Noticing that $\gamma_i$ for $i \in \{ 1,2\}$ in (\ref{LN:1}) are gamma random variables, the desired result can be obtained by following the similar lines as in the proof of Theorem \ref{theorem:3}.
\endproof
\end{theorem}

{Recall the exact ergodic capacity of ZF/MRT scheme in (20), when $N$ is sufficiently large, for a fixed $M$, we have $N-M\approx N$, hence (20) reduces to (26), which confirms the correctness of Theorem \ref{theorem:6}. In addition, Theorem \ref{theorem:6} can be also viewed as the exact ergodic capacity of dual-hop AF relaying systems operating over Nakagami-m fading channels. Hence, it extends the analysis of \cite{FAN}, which deals with the Rayleigh fading channels.}

\section{Numerical Results and Discussion}\label{section:4}
In this section, we present numerical results to validate the analytical expressions derived in Section III. Unless otherwise stated, we set $\rho_1 = \rho_2$, i.e., a symmetric setting where the relay is spaced equal-distant from the source and the destination and all Monte Carlo simulation results are obtained with $10^5$ runs.

\begin{figure}[ht]
\centering
\includegraphics[width=0.4\textwidth]{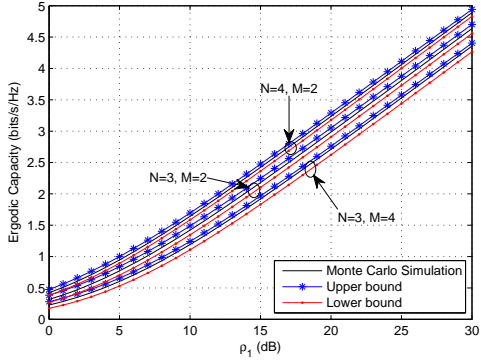}
\caption{Ergodic capacity of the MRC/MRT scheme with different $N$, $M$.}\label{fig:fig2}
\end{figure}

Fig. \ref{fig:fig2} examines the ergodic capacity of the MRC/MRT scheme with different $N$ and $M$. As shown in the figure, for all simulation setups, the proposed upper bound and lower bound are sufficiently tight across the entire SNR range of interest. It is also evident that the increasing $N$ improves the ergodic capacity performance of the system. Moreover, we observe the intuitive result that increasing $M$ results in a degradation of the ergodic capacity performance of the system. In addition, we see that the tightness of the proposed lower and upper bounds improve as $N$ grows large.

\begin{figure}[ht]
\centering
\includegraphics[width=0.4\textwidth]{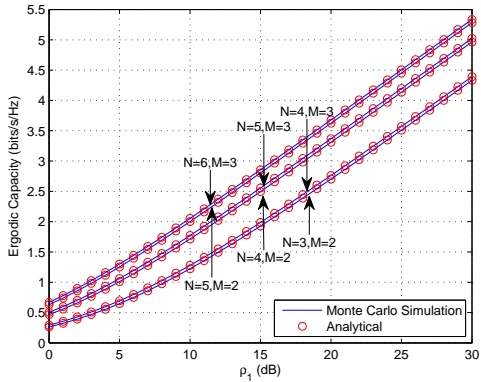}
\caption{Ergodic capacity of the ZF/MRT scheme with different $N$ and $M$.}\label{fig:fig3}
\end{figure}

Fig. \ref{fig:fig3} illustrates the ergodic capacity of the ZF/MRC scheme with different $N$ and $M$.  We see that the analytical results in Theorem \ref{theorem:3} are in exact agreement with the Monte Carlo simulation results, hence confirming the correctness of the analytical expression. Again, it is observed that, for fixed $M$, increasing the antenna number $N$ yields a significant capacity improvement. {Moreover, we observe that, for a fixed $N-M$, the ergodic capacity difference between different $M$, $N$ pairs is almost negligible.}

\begin{figure}[ht]
\centering
\includegraphics[width=0.4\textwidth]{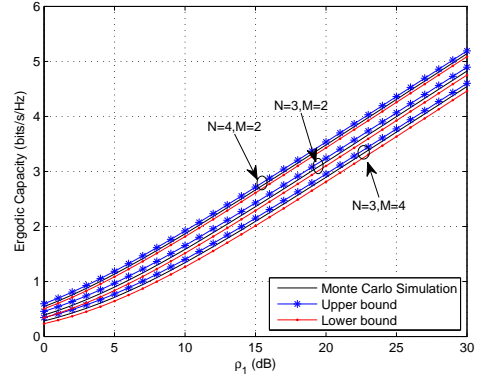}
\caption{Ergodic capacity of the MMSE/MRT scheme with different $N$ and $M$.}\label{fig:fig4}
\end{figure}
Fig. \ref{fig:fig4} shows the ergodic capacity of the MMSE/MRT scheme with different $N$ and $M$. We can readily note that both the upper bound and the lower bound remains sufficiently tight across the entire SNR range of interest, which means both of them are able to serve as an effective approximation to the exact ergodic capacity value. In addition, we see that the impact of $N$ and $M$ on the ergodic capacity is similar to that of the MRC/MRT scheme.

\begin{figure}[ht]
\centering
\includegraphics[width=0.4\textwidth]{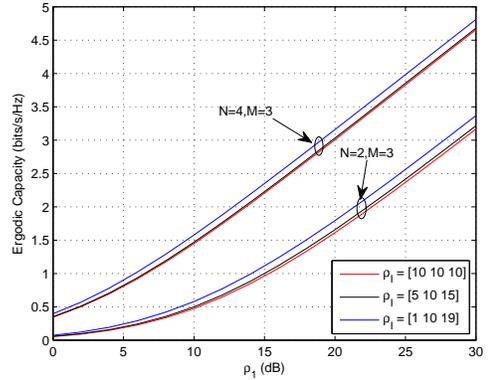}
\caption{The ergodic capacity of the MMSE/MRT scheme with different interference power distributions and $N$.}\label{fig:fig5}
\end{figure}
Fig. \ref{fig:fig5} examines the effect of interference power distribution on the ergodic capacity of the MMSE/MRT scheme. Two sets of curves are plotted. As we can readily observe, for a given total interference power, {the ergodic capacity of the system subject to equal-power interferers appears as a tight lower bound for the scenario with unequal-power interferers}. Moreover, the performance gaps among them becomes closer as $N$ grows large.  This observation also implies that, with the MMSE/MRT scheme, for a given total received interference power, an equal interference power scenario yields the worst ergodic capacity performance.

\begin{figure}[ht]
\centering
\includegraphics[width=0.4\textwidth]{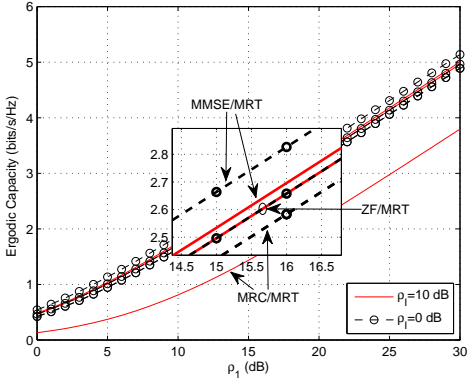}
\caption{Ergodic capacity comparison among the MRC/MRT, ZF/MRT and MMSE/MRT schemes with $N = 4$, $M = 2$.}\label{fig:fig6}
\end{figure}
Fig. \ref{fig:fig6} compares the ergodic capacity of the three linear processing schemes under different interference power,  i.e., weak interference $\rho_I=0 \mbox{ dB}$ and strong interference $\rho_I=10\mbox{ dB}$. It can be easily observed that, in both cases, the MMSE/MRT scheme always has the best performance and the ZF/MRT scheme is slightly inferior, while the MRC/MRT scheme is always the worst one. Moreover, when the interference power is small, i.e., $\rho_I = 0 \mbox{ dB}$, the capacity difference of three schemes is quite small. However, as the interference power grows large, i.e., $\rho_I = 10 \mbox{ dB}$, the capacity gap between the MMSE/MRT scheme and the ZF/MRT scheme narrows down, while the difference between the MMSE/MRT scheme and the MRC/MRT scheme increases significantly. This observation suggests that, in the presence of weak interference, the MRC/MRT scheme may be a good choice in practice because of its low implementation complexity. However, when the interference is strong, more sophisticated schemes with superior interference suppression capability, i.e., the ZF/MRT or the MMSE/MRT scheme should be used.

\begin{figure}[ht]
\centering
\includegraphics[width=0.4\textwidth]{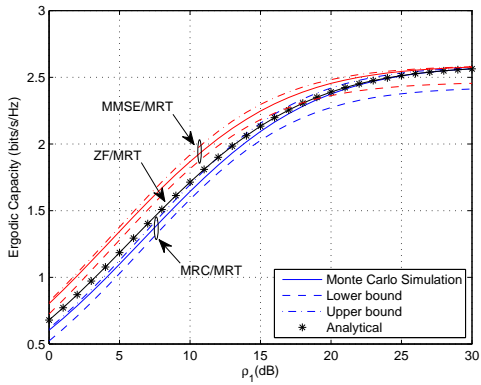}
\caption{Ergodic capacity comparison among the MRC/MRT, ZF/MRT and MMSE/MRT schemes with fixed $\rho_2=10\;{\mbox{dB}}$ and $N = 4$, $M = 2$, $\rho_I=0\;{\mbox{dB}}$}\label{fig:fig8}
\end{figure}
{Fig.\ref{fig:fig8} compares the ergodic capacity of the proposed three schemes with fixed $\rho_2=10\;{\mbox{dB}}$. We can readily note that, the proposed upper and the lower bounds remain sufficiently tight across the entire range of SNRs of interest. In addition, the upper bounds become almost exact in the high SNR regime. Moreover, we see that fixing $\rho_2$ results in the ``ceiling effect'' for all three schemes, which is rather intuitive since the capacity of dual-hop systems is limited by the quality of the weakest hop. Finally, we observe that, when $\rho_1$ is large, the performance gap among the three schemes becomes negligible. The underlying reason is that, as $\rho_1$ grows large, the strength of the desired signal improves considerably, hence the advantage of the MMSE/MRT and ZF/MRT schemes in terms of interference suppression becomes less pronounced, and with a fixed $\rho_2$, the quality of the second hop is the bottleneck, which is the same for all three schemes.}


\begin{figure}[ht]
\centering
\includegraphics[width=0.4\textwidth]{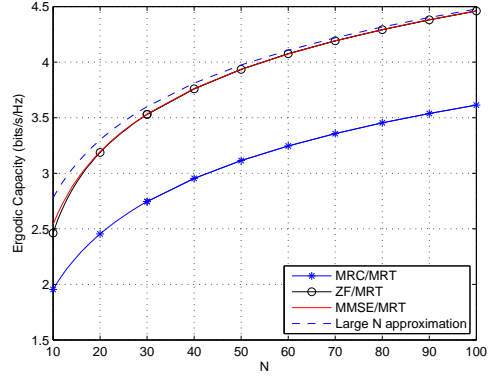}
\caption{Ergodic capacity: Large $N$ regime with $\rho_1=\rho_2=10\;{\mbox{dB}}$, $\rho_I=0\;{\mbox{dB}}$ and $M=5$.}\label{fig:fig7}
\end{figure}
Fig. \ref{fig:fig7} investigates the impact of $N$ on the ergodic capacity performance of three proposed schemes. As expected, the ergodic capacity of all the three schemes increases as $N$ becomes large. Moreover, the rate of increasing gradually becomes smaller. In addition, we observe that the ZF/MRT and the MMSE/MRT schemes attain the same capacity when $N$ is sufficiently large, i.e., $N\geq 20$. However, there is a significant gap between the MMSE/MRT scheme and the MRC/MRT scheme, and such gap does not seem to diminish as $N$ grows large, instead, it remains more or less unchanged. These important observations suggest that, in the large $N$ regime, both the ZF/MRT and the MMSE/MRT schemes are capable of perfect interference cancelation, which is not possible with the MRC/MRT scheme.

\section{Conclusions}
In this paper, we have investigated the ergodic capacity of the MRC/MRT, ZF/MRT and MMSE/MRT schemes in an AF relaying system with CCI at the multiple antenna relay node. New analytical exact or tight upper/lower bounds were derived for the ergodic capacity, which not only provide an efficient means for the evaluation of the ergodic capacity, but also enable the characterization of the impact of key system parameters such as antenna number $N$, CCI number $M$ and interference power on the performance of the system. Our findings suggest that, the MMSE/MRT scheme always attains the highest capacity and the ZF/MRT scheme is slightly inferior, while the MRC/MRT scheme is always the worst one. Moreover, in the large $N$ regime, both the ZF/MRT and MMSE/MRT schemes have perfect interference cancelation capability, which is not possible with the MRC/MRT scheme.

\appendices
\section{Proof for the MRC/MRT Scheme}
\subsection{Proof of Theorem \ref{theorem:1}}\label{appendix:theorem:1}
Combining (\ref{C:4}) and (\ref{C:5}), the ergodic capacity of the MRC/MRT scheme can be upper bounded by
\begin{multline}\label{MRC:a19}
C_{\sf MRC}^{\sf up} = C_{\gamma _1^{\sf MRC}}+C_{\gamma _2^{\sf MRC}}\\
-\frac{1}{2}{\log _2}\left( {1 + {e^{{\mathop{\rm E}\nolimits} \left( {\ln {\gamma _1^{\sf MRC}}} \right)}} + {e^{{\mathop{\rm E}\nolimits} \left( {\ln {\gamma _2^{\sf MRC}}} \right)}}} \right).
\end{multline}
We now evaluate the four items $C_{\gamma _1^{\sf MRC}}$, $C_{\gamma _2^{\sf MRC}}$, ${{\mathop{\rm E}\nolimits} \left( {\ln {\gamma _1^{\sf MRC}}} \right)}$ and ${{\mathop{\rm E}\nolimits} \left( {\ln {\gamma _2^{\sf MRC}}} \right)}$ in the following part.

\subsubsection{Calculation of $C_{\gamma _1^{\sf MRC}}$}
We first note that $C_{\gamma _1^{\sf MRC}}$ can be computed by \cite{Himal22}
\begin{align}\label{MRC:a1}
{C_{{\gamma _1^{\sf MRC}}}} = \frac{1}{{2\ln 2}}\int_0^\infty  {\frac{{1 - {F_{{\gamma _1^{\sf MRC}}}}\left( x \right)}}{{1 + x}}dx}.
\end{align}
Then, invoking the c.d.f. of $\gamma _1^{\sf MRC}$ \cite{ZHU_TWC}
\begin{multline}\label{MRC:a2}
{F _{{\gamma _1^{\sf MRC}}}}\left( x \right) = 1 - {e^{ - \frac{x}{{{\rho _1}}}}}\sum\limits_{k = 0}^{N - 1} {\frac{{{x^k}}}{{\rho _1^kk!}}\sum\limits_{l = 0}^k {{k\choose l}\sum\limits_{i = 1}^{\rho \left( {\bf{D}} \right)} {\sum\limits_{j = 1}^{{\tau _i}({\bf{D}})} {{\chi _{i,j}}({\bf{D}})} } } }\\
\times\frac{{\Gamma \left( {j + l} \right)}}{{\Gamma \left( j \right)}}\rho _{I\left\langle i \right\rangle }^l{{\left( {\frac{{{\rho _1}}}{{{\rho _1} + {\rho _{I\left\langle i \right\rangle }}x}}} \right)}^{j + l}},
\end{multline}
the integral in (\ref{MRC:a1}) can be evaluated as
\begin{multline}\label{MRC:a3}
{C_{{\gamma _1^{\sf MRC}}}} =\\ \frac{1}{{2\ln 2}}\sum\limits_{k = 0}^{N - 1} {\frac{1}{{\rho _1^kk!}}\sum\limits_{l = 0}^k {k\choose l} } \sum\limits_{i = 1}^{\rho ({\bf{D}})} {\sum\limits_{j = 1}^{{\tau _i}({\bf{D}})} {{\chi _{i,j}}({\bf{D}})\frac{{\Gamma \left( {j + l} \right)}}{{\Gamma \left( j \right)}}}}\\
 \times\rho _{I\left\langle i \right\rangle }^l  \underbrace {\int_0^\infty  {{e^{ - \frac{x}{{{\rho _1}}}}}{x^k}{{\left( {1 + x} \right)}^{ - 1}}{{\left( {\frac{{{\rho _1}}}{{{\rho _1} + {\rho _{I\left\langle i \right\rangle }}x}}} \right)}^{j + l}}dx} }_{{{\cal I}_1}}.
\end{multline}
To this end, noticing that ${\left( {1 + \beta x} \right)^{ - \alpha }} = \frac{1}{{\Gamma \left( \alpha  \right)}}{\mathop{\rm G}\nolimits} _{1,1}^{1,1}\left( \beta x\middle| \substack{{1 - \alpha }\\
0} \right)$,
and with the help of the formula \cite[Eq. (2.6.2)]{H-function}, we obtain
\begin{align}\label{MRC:a4}
{{\cal I}_1} = \frac{{\rho _1^{k + 1}}}{{\Gamma \left( {j + l} \right)}}{\mathop{\rm G}\nolimits} _{1,[1:1],0,[1:1]}^{1,1,1,1,1}\left(^{{\rho _1}}_{{\rho _{I\left\langle i \right\rangle }}}
\middle| \substack{
{k + 1}\\
{0;1 - j - l}\\
 - \\
{0;0}} \right).
\end{align}
Finally, substituting (\ref{MRC:a4}) into (\ref{MRC:a3}) ${C_{{\gamma _1^{\sf MRC}}}}$ can be expressed in compact-form as
\begin{multline}\label{MRC:a16}
{C_{{\gamma _1^{\sf MRC}}}} = \frac{{{\rho _1}}}{{2\ln 2}}\sum\limits_{k = 0}^{N - 1} {\frac{1}{{k!}}\sum\limits_{l = 0}^k {k\choose l} } \sum\limits_{i = 1}^{\rho ({\bf{D}})} {\sum\limits_{j = 1}^{{\tau _i}({\bf{D}})} {{\chi _{i,j}}({\bf{D}})\frac{{\rho _{I\left\langle i \right\rangle }^l}}{{\Gamma \left( j \right)}}}}\\
\times{\mathop{\rm G}\nolimits} _{1,[1:1],0,[1:1]}^{1,1,1,1,1}\left(^{{\rho _1}}_{{\rho _{I\left\langle i \right\rangle }}}
\middle| \substack{
{k + 1}\\
{0;1 - j - l}\\
 - \\
{0;0}} \right).
\end{multline}

\subsubsection{Calculation of $C_{\gamma _2^{\sf MRC}}$}
Similarly, $C_{\gamma _2^{\sf MRC}}$ can be computed by
\begin{align}\label{MRC:a5}
{C_{{\gamma _2^{\sf MRC}}}} = \frac{1}{{2\ln 2}}\int_0^\infty  {\frac{{1 - {F_{{\gamma _2^{\sf MRC}}}}\left( x \right)}}{{1 + x}}dx}.
\end{align}
Noticing that $\gamma _2^{\sf MRC}$ is a gamma random variable with the c.d.f. given by
\begin{align}\label{MRC:a6}
{F _{{\gamma _2^{\sf MRC}}}}\left( x \right) = 1 - {e^{ - \frac{x}{{{\rho _2}}}}}\sum\limits_{m = 0}^{N - 1} {\frac{{{x^m}}}{{\rho _2^mm!}}}.
\end{align}
Now (\ref{MRC:a5}) can be written as
\begin{align}
{C_{{\gamma _2^{\sf MRC}}}} = \frac{1}{{2\ln 2}}\sum\limits_{k = 0}^{N - 1} {\frac{1}{{k!}}} {\left( {\frac{1}{{{\rho _2}}}} \right)^k}\int_0^\infty  {\frac{{{e^{\frac{x}{{{\rho _2}}}}}{x^k}}}{{1 + x}}dx}.
\end{align}
Finally, utilizing \cite[Eq. (3.383.10)]{Tables}, ${C_{{\gamma _2^{\sf MRC}}}}$ can be expressed in closed-form as
\begin{align}\label{MRC:a17}
{C_{{\gamma _2^{\sf MRC}}}} = \frac{1}{{2\ln 2}}{e^{\frac{1}{{{\rho _2}}}}}\sum\limits_{k = 0}^{N - 1} {{{\left( {\frac{1}{{{\rho _2}}}} \right)}^k}\Gamma \left( { - k,\frac{1}{{{\rho _2}}}} \right)}.
\end{align}

\subsubsection{Calculation of ${{\mathop{\rm E}\nolimits} \left( {\ln {\gamma _1^{\sf MRC}}} \right)}$}
The expectation of $\ln {\gamma _1^{\sf MRC}}$ can be derived from
\begin{align}\label{MRC:a7}
{\mathop{\rm E}\nolimits} \left( {\ln \gamma _1^{\sf MRC}} \right) = {\left. {\frac{{d{\mathop{\rm E}\nolimits} \left( {{{\left( {\gamma _1^{\sf MRC}} \right)}^n}} \right)}}{{dn}}} \right|_{n = 0}},
\end{align}
where we have used the following derivative property
\begin{align}
\frac{{d{x^n}}}{{dn}} = {x^n}\ln x.
\end{align}
Hence, the first step is to work out the general moment of ${\gamma _1^{\sf MRC}}$. For a non-negative random variable $X$, its general moment can be computed via
\begin{align}\label{MRC:a8}
{\rm{E}}\left( {{x^n}} \right) = n\int_0^\infty  {{x^{n - 1}}\left( {1 - {F _X}\left( x \right)} \right)dx},
\end{align}
where $F_X(x)$ is the c.d.f. of $X$.
Hence, we have
\begin{multline}\label{MRC:a9}
{\rm{E}}\left( {{{\left( {\gamma _1^{\sf MRC}} \right)}^n}} \right) = \sum\limits_{k = 0}^{N - 1} {\frac{1}{{\rho _1^kk!}}\sum\limits_{l = 0}^k {k\choose l} } \sum\limits_{i = 1}^{\rho ({\bf{D}})} {\sum\limits_{j = 1}^{{\tau _i}({\bf{D}})} {{\chi _{i,j}}({\bf{D}})} }\\ \times \frac{{\Gamma \left( {j + l} \right)}}{{\Gamma \left( j \right)}}
\rho _{I\left\langle i \right\rangle }^l n{{\cal I}_2},
\end{multline}
where ${{\cal I}_2}= {\int_0^\infty  {{e^{ - \frac{x}{{{\rho _1}}}}}{x^{k + n - 1}}{{\left( {1 + \frac{{{\rho _{I\left\langle i \right\rangle }}}}{{{\rho _1}}}} \right)}^{ - (j + l)}}dx} }$.
Invoking \cite[Eq. (9.211.4)]{Tables}, (\ref{MRC:a9}) can be alternatively given by

\begin{multline}\label{MRC:a25}
{\rm{E}}\left( {{{\left( {\gamma _1^{\sf MRC}} \right)}^n}} \right) = \sum\limits_{k = 0}^{N - 1} {\frac{1}{{\rho _1^kk!}}\sum\limits_{l = 0}^k {k\choose l} } \times\\\sum\limits_{i = 1}^{\rho ({\bf{D}})} {\sum\limits_{j = 1}^{{\tau _i}({\bf{D}})} {{\chi _{i,j}}({\bf{D}})\frac{{\Gamma \left( {j + l} \right)}}{{\Gamma \left( j \right)}}\rho _{I\left\langle i \right\rangle }^l} } n{\left( {\frac{{{\rho _1}}}{{{\rho _{I\left\langle i \right\rangle }}}}} \right)^{k + n}}\times\\\Gamma \left( {k + n} \right)\Psi \left( {k + n,k + n - j - l + 1;\frac{1}{{{\rho _{I\left\langle i \right\rangle }}}}} \right).
\end{multline}

To proceed with the computation, it is convenient to use the alternative expression as (\ref{MRC:a25.1}) shown on the top of the next page,
\begin{figure*}
\begin{multline}\label{MRC:a25.1}
{\rm{E}}\left( {{{\left( {\gamma _1^{\sf MRC}} \right)}^n}} \right) = \underbrace {\sum\limits_{i = 1}^{\rho ({\bf{D}})} {\sum\limits_{j = 1}^{{\tau _i}({\bf{D}})} {{\chi _{i,j}}({\bf{D}})} } {{\left( {\frac{{{\rho _1}}}{{{\rho _{I\left\langle i \right\rangle }}}}} \right)}^n}\Gamma \left( {n + 1} \right)\Psi \left( {n,n - j + 1;\frac{1}{{{\rho _{I\left\langle i \right\rangle }}}}} \right)}_{{s_1}\left( n \right)} + \\
\underbrace {\sum\limits_{k = 1}^{N - 1} {\frac{1}{{\rho _1^kk!}}\sum\limits_{l = 0}^k {k\choose l} } \sum\limits_{i = 1}^{\rho ({\bf{D}})} {\sum\limits_{j = 1}^{{\tau _i}({\bf{D}})} {{\chi _{i,j}}({\bf{D}})\frac{{\Gamma \left( {j + l} \right)}}{{\Gamma \left( j \right)}}\rho _{I\left\langle i \right\rangle }^ln{T_1}\left( n \right)} } }_{{s_2}\left( n \right)},
\end{multline}
\hrule
\end{figure*}
where $ {T_1}\left( n \right) = {\left( {\frac{{{\rho _1}}}{{{\rho _{I\left\langle i \right\rangle }}}}} \right)^{k + n}}\Gamma \left( {k + n} \right)
\Psi \left( {k + n,k + n - j - l + 1;\frac{1}{{{\rho _{I\left\langle i \right\rangle }}}}} \right)$.
Then, according to (\ref{MRC:a7}), the expectation of $\ln {\gamma _1^{\sf MRC}}$  can be computed as
\begin{align}\label{MRC:a11}
{\mathop{\rm E}\nolimits} \left( {\ln {\gamma _1^{\sf MRC}}} \right) = {\left. {\frac{{d{s_1}\left( n \right)}}{{dn}}} \right|_{n = 0}} + {\left. {\frac{{d{s_2}\left( n \right)}}{{dn}}} \right|_{n = 0}}.
\end{align}
We start with the computation of ${\left. {\frac{{d{s_1}\left( n \right)}}{{dn}}} \right|_{n = 0}}$, and it is easy to have (\ref{MRC:a12}) shown on the top of the next page,
\begin{multline}\label{MRC:a12}
{\left. {\frac{{d{s_1}\left( n \right)}}{{dn}}} \right|_{n = 0}} = \sum\limits_{i = 1}^{\rho ({\bf{D}})} {\sum\limits_{j = 1}^{{\tau _i}({\bf{D}})} {{\chi _{i,j}}({\bf{D}})}}\left[ {\left( {\ln \frac{{{\rho _1}}}{{{\rho _{I\left\langle i \right\rangle }}}} + \psi \left( 1 \right)} \right)} \right.\\ \times \Psi \left( {0,1 - j;\frac{1}{{{\rho _{I\left\langle i \right\rangle }}}}} \right) +
{\Psi ^{\left( {1,0,0} \right)}}\left( {0,1 - j;\frac{1}{{{\rho _{I\left\langle i \right\rangle }}}}} \right) \\+ \left. {{\Psi ^{\left( {0,1,0} \right)}}\left( {0,1 - j;\frac{1}{{{\rho _{I\left\langle i \right\rangle }}}}} \right)} \right].
\end{multline}

To compute ${\left. {\frac{{d{s_2}\left( n \right)}}{{dn}}} \right|_{n = 0}}$, we observe that the key task is to compute ${\left. {\frac{{dn{T_1}\left( n \right)}}{{dn}}} \right|_{n = 0}}$, and we have
\begin{align}
{\left. {\frac{{dn{T_1}\left( n \right)}}{{dn}}} \right|_{n = 0}} = {\left. {{T_1}\left( n \right)} \right|_{n = 0}} + n{\left. {\frac{{d{T_1}\left( n \right)}}{{dn}}} \right|_{n = 0}}.
\end{align}
Noticing that, when $k \ge 1$, ${\left. {\frac{{d{T_1}\left( n \right)}}{{dn}}} \right|_{n = 0}} < \infty $ is a constant, hence $n{\left. {\frac{{d{T_1}\left( n \right)}}{{dn}}} \right|_{n = 0}} = 0$. Then, we obtain
\begin{multline}\label{MRC:a13}
{\left. {\frac{{d{s_2}\left( n \right)}}{{dn}}} \right|_{n = 0}} = \sum\limits_{k = 1}^{N - 1} {\frac{1}{k}\sum\limits_{l = 0}^k {k\choose l} } \sum\limits_{i = 1}^{\rho ({\bf{D}})} {\sum\limits_{j = 1}^{{\tau _i}({\bf{D}})} {{\chi _{i,j}}({\bf{D}})} }\\
\times\frac{{\Gamma \left( {j + l} \right)}}{{\Gamma \left( j \right)}}\rho _{I\left\langle i \right\rangle }^{l - k}\Psi \left( {k,k - j - l + 1;\frac{1}{{{\rho _{I\left\langle i \right\rangle }}}}} \right).
\end{multline}
To this end, substituting (\ref{MRC:a12}) and (\ref{MRC:a13}) into (\ref{MRC:a11}),  the expectation of $\ln {\gamma _1^{\sf MRC}}$  can be expressed as (\ref{MRC:a14}) shown on the top of the next page.
\begin{figure*}
\begin{multline}\label{MRC:a14}
{\mathop{\rm E}\nolimits} \left( {\ln {\gamma _1^{\sf MRC}}} \right) = \sum\limits_{i = 1}^{\rho ({\bf{D}})} {\sum\limits_{j = 1}^{{\tau _i}({\bf{D}})} {{\chi _{i,j}}({\bf{D}})\left[ {\left( {\ln \frac{{{\rho _1}}}{{{\rho _{I\left\langle i \right\rangle }}}} + \psi \left( 1 \right)} \right)\Psi \left( {0,1 - j;\frac{1}{{{\rho _{I\left\langle i \right\rangle }}}}} \right)} \right.} } + {\Psi ^{\left( {1,0,0} \right)}}\left( {0,1 - j;\frac{1}{{{\rho _{I\left\langle i \right\rangle }}}}} \right) + \\\left. {{\Psi ^{\left( {0,1,0} \right)}}\left( {0,1 - j;\frac{1}{{{\rho _{I\left\langle i \right\rangle }}}}} \right)} \right]
 + \sum\limits_{k = 1}^{N - 1} {\frac{1}{k}\sum\limits_{l = 0}^k {k\choose l} } \sum\limits_{i = 1}^{\rho ({\bf{D}})} {\sum\limits_{j = 1}^{{\tau _i}({\bf{D}})} {{\chi _{i,j}}({\bf{D}})\frac{{\Gamma \left( {j + l} \right)}}{{\Gamma \left( j \right)}}\rho _{I\left\langle i \right\rangle }^{l - k}\Psi \left( {k,k - j - l + 1;\frac{1}{{{\rho _{I\left\langle i \right\rangle }}}}} \right)} }.
\end{multline}
\hrule
\end{figure*}

\subsubsection{Calculation of  ${{\mathop{\rm E}\nolimits} \left( {\ln {\gamma _2^{\sf MRC}}} \right)}$}
Since ${\gamma _2^{\sf MRC}}$ is a gamma random variable, the expectation of $\ln {\gamma _2^{\sf MRC}}$ can be derived directly as
\begin{align}
{\mathop{\rm E}\nolimits} \left( {\ln {\gamma _2^{\sf MRC}}} \right) = {\left( {\frac{1}{{{\rho _2}}}} \right)^N}\frac{1}{{\Gamma \left( N \right)}}\int_0^\infty  {{x^{N - 1}}{e^{ - \frac{x}{{{\rho _2}}}}}\ln xdx}.
\end{align}
Utilizing \cite[Eq. (4.352.1)]{Tables}, we obtain
\begin{align}\label{MRC:a18}
{\mathop{\rm E}\nolimits} \left( {\ln {\gamma _2^{\sf MRC}}} \right) = \psi \left( N \right) + \ln {\rho _2}.
\end{align}
Finally, substituting (\ref{MRC:a16}), (\ref{MRC:a17}), (\ref{MRC:a14}) and (\ref{MRC:a18}) into (\ref{MRC:a19}) yields the desired result.

\subsection{Proof of Theorem \ref{theorem:2}}\label{appendix:theorem:2}
Combining (\ref{C:4}) and (\ref{C:6}), the ergodic capacity lower bound of the MRC/MRT scheme can be computed as
\begin{multline}\label{MRC:a20}
{C_{\sf MRC}^{\sf low}} = {C_{{\gamma _1^{\sf MRC}}}} + {C_{{\gamma _2^{\sf MRC}}}} \\- \frac{1}{2}{\log _2}\left( {1 + {\mathop{\rm E}\nolimits} \left( {{\gamma _1^{\sf MRC}}} \right) + {\mathop{\rm E}\nolimits} \left( {{\gamma _2^{\sf MRC}}} \right)} \right).
\end{multline}
Since ${C_{{\gamma _1^{\sf MRC}}}}$ and ${C_{{\gamma _2^{\sf MRC}}}}$ have been derived in (\ref{MRC:a16}) and (\ref{MRC:a17}) respectively. The remaining task is figure out ${\mathop{\rm E}\nolimits} \left( {{\gamma _1^{\sf MRC}}} \right)$ and ${\mathop{\rm E}\nolimits} \left( {{\gamma _2^{\sf MRC}}} \right)$.

\subsubsection{Calculation of ${\mathop{\rm E}\nolimits} \left( {{\gamma _1^{\sf MRC}}} \right)$}
Setting $n=1$ in (\ref{MRC:a25}), we get
\begin{multline}\label{MRC:a23}
{\mathop{\rm E}\nolimits} \left( {{\gamma _1^{\sf MRC}}} \right) = {\rho _1}\sum\limits_{k = 0}^{N - 1} {\sum\limits_{l = 0}^k {k\choose l} } \sum\limits_{i = 1}^{\rho ({\bf{D}})} {\sum\limits_{j = 1}^{{\tau _i}({\bf{D}})} {{\chi _{i,j}}({\bf{D}})} }\\ \times\frac{{\Gamma \left( {j + l} \right)}}{{\Gamma \left( j \right)}}\rho _{I\left\langle i \right\rangle }^{l - k - 1}\Psi \left( {k + 1,k - j - l + 2;\frac{1}{{{\rho _{I\left\langle i \right\rangle }}}}} \right).
\end{multline}

\subsubsection{Calculation of ${\mathop{\rm E}\nolimits} \left( {{\gamma _2^{\sf MRC}}} \right)$}
With the help of (\ref{MRC:a8}), the expectation of ${\gamma _2^{\sf MRC}}$ can be computed as
\begin{align}\label{MRC:a24}
{\mathop{\rm E}\nolimits} \left( {{\gamma _2^{\sf MRC}}} \right) = \sum\limits_{m = 0}^{N - 1} {\frac{1}{{m!}}\int_0^\infty  {{e^{ - \frac{x}{{{\rho _2}}}}}{{\left( {\frac{x}{{{\rho _2}}}} \right)}^m}dx} } = N{\rho _2}.
\end{align}
Finally, Substituting (\ref{MRC:a16}), (\ref{MRC:a17}), (\ref{MRC:a23}) and (\ref{MRC:a24}) into (\ref{MRC:a20}) yields the desired result.

\section{Proof for the ZF/MRT Scheme}
\subsection{Proof of Theorem \ref{theorem:3}}\label{appendix:theorem:3}
Substituting  (\ref{C:7}) into (\ref{C:1}), the ergodic capacity of the ZF/MRT scheme is given by
\begin{align}\label{ZF:c1}
C_{\sf ZF}= C_{{\gamma _1^{\sf ZF}}}+C_{{\gamma _2^{\sf ZF}}}- C_{{\gamma _T^{\sf ZF}}},
\end{align}
where $C_{{\gamma _i^{\sf ZF}}} = \frac{1}{2}{{\mathop{\rm E}\nolimits} }\left[ {{{\log }_2}\left( {1 + {\gamma _i^{\sf ZF}}} \right)} \right]$ for $i \in \{ 1,2\}$ and $C_{{\gamma _T^{\sf ZF}}} = \frac{1}{2}{{\mathop{\rm E}\nolimits} }\left[ {{{\log }_2}\left( {1 + {\gamma _1^{\sf ZF}}+{\gamma _2^{\sf ZF}}} \right)} \right]$.

To proceed, we need to find out the statistics of $\gamma_i^{\sf ZF}$ for $i \in \{ 1,2\}$. From \cite{Z.Ding}, the probability density function (p.d.f.) of $y_1={\left| {{\bf{h}}_1^\dag {\bf{P}}{{\bf{h}}_1}} \right|}$ is given by ${f_{{y_1}}}\left( x \right) = \frac{{{x^{N - M - 1}}}}{{\left( {N - M - 1} \right)!}}{e^{ - x}}$,
and we know
${f_{{{{\| {{{\bf{h}}_{2}}} \|}^2_F}}}}(x) = \frac{{{x^{N - 1}}}}{{(N - 1)!}}{e^{ - x}}$. Then, using \cite[Eq. (8.352.4)]{Tables}, the c.d.f. of $\gamma_i^{\sf ZF}$ for $i \in \{ 1,2\}$ can be written as
\begin{align}\label{ZF:c2}
{F_{{\gamma _i^{\sf ZF}}}}\left( x \right) = 1 - \frac{{\Gamma \left( {{N_i},\frac{x}{{{\rho _1}}}} \right)}}{{\Gamma \left( {{N_i}} \right)}} = 1 - {e^{ - \frac{x}{{{\rho _i}}}}}\sum\limits_{k = 0}^{{N_i} - 1} {\frac{1}{{k!}{\rho _i^k}}}{x^k},
\end{align}
where $N_1 = N-M$ and $N_2 = N$.


\subsubsection{Calculation of ${C_{{\gamma _i^{\sf ZF}}}}$}
Similar to (\ref{MRC:a1}), and invoking (\ref{ZF:c2}) and \cite[Eq. (8.383.10)]{Tables},  we have
\begin{align}\label{ZF:c10}
{C_{{\gamma _i^{\sf ZF}}}} &=\frac{1}{{2\ln 2}}\sum\limits_{k = 0}^{{N_i}-1} {\frac{1}{{k!}}} {\left( {\frac{1}{{{\rho _i}}}} \right)^k}\int_0^\infty  {\frac{{{e^{\frac{x}{{{\rho _i}}}}}{x^k}}}{{1 + x}}dx} \notag \\
&=\frac{1}{{2\ln 2}}\sum\limits_{k = 0}^{{N_i} - 1} {\left( {\frac{1}{{{\rho _i}}}} \right)^k}{e^{\frac{1}{{{\rho _i}}}}}\Gamma \left( { - k,\frac{1}{{{\rho _i}}}} \right).
\end{align}

\subsubsection{Calculation of  ${C_{{\gamma _T^{\sf ZF}}}}$}
Since the c.d.f. expression of $\gamma_T^{\sf ZF}$ is in general difficult to obtain, the above c.d.f. based approach can not be applied here. Instead, we adopt an alternative moment generating function (MGF) based approach \cite{K.A.Hamdi} to compute ${C_{{\gamma _T^{\sf ZF}}}}$.
\begin{align}\label{ZF:c4}
{C_{{\gamma _T^{\sf ZF}}}} = \frac{1}{{2\ln 2}}\int_0^\infty  {\frac{{{e^{ - s}}}}{s}\left( {1 - {{\mathop{\rm M}\nolimits} _{{\gamma _T^{\sf ZF}}}}\left( s \right)} \right)ds},
\end{align}
where ${{\mathop{\rm M}\nolimits} _{{\gamma _T^{\sf ZF}}}}\left( s \right)$ is the MGF of $\gamma_T^{\sf ZF}$.

The MGF of $\gamma_i^{\sf ZF}$ can be computed by
\begin{align}\notag
{{\mathop{\rm M}\nolimits} _{{\gamma _i^{\sf ZF}}}}\left( s \right) = \frac{\int_0^\infty  {{x^{{N_i-1}}}{e^{ - \left( {s + \frac{1}{{{\rho _i}}}} \right)x}}dx}}{{\Gamma \left( {{N_i}} \right)\rho _i^{{N_i}}}} = {\left( {\frac{1}{{1 + s{\rho _i}}}} \right)^{{N_i}}}\notag.
\end{align}
Since $\gamma_1^{\sf ZF}$ and $\gamma_2^{\sf ZF}$ are independent random variables, we have
\begin{align}\notag
{{\mathop{\rm M}\nolimits} _{{\gamma _T^{\sf ZF}}}}( s )& = {{\mathop{\rm M}\nolimits} _{{\gamma _1^{\sf ZF}}}}( s ){{\mathop{\rm M}\nolimits} _{{\gamma _2^{\sf ZF}}}}( s )
={\left( {\frac{1}{{1 + s{\rho _1}}}} \right)^{{N_1}}}{\left( {\frac{1}{{1 + s{\rho _2}}}} \right)^{{N_2}}}.
\end{align}
To avoid the singularity problem caused by the term ${\frac{{{e^{ - s}}}}{s}}$ around zero when evaluating  (\ref{ZF:c4}), we find it convenient to use the following alternative MGF expression for $\gamma_i^{\sf ZF}$
\begin{align}
\begin{array}{l}
{{\mathop{\rm M}\nolimits} _{{\gamma _i^{\sf ZF}}}}\left( s \right) = 1 - s{\rho _i}\sum\limits_{k = 0}^{{N_i} - 1} {{{\left( {\frac{1}{{1 + s{\rho _i}}}} \right)}^{k + 1}}}.
\end{array}
\end{align}
%
%
Then, the MGF of ${\gamma _T^{\sf ZF}}$ can be alternatively expressed as
\begin{multline}\label{ZF:c5}
{{\mathop{\rm M}\nolimits} _{{\gamma _T^{\sf ZF}}}}\left( s \right) = \\1 - s{\rho _1}\sum\limits_{k = 0}^{N - M - 1} {{{\left( {\frac{1}{{1 + s{\rho _1}}}} \right)}^{k + 1}}}  - s{\rho _2}\sum\limits_{j = 0}^{N - 1} {{{\left( {\frac{1}{{1 + s{\rho _2}}}} \right)}^{j + 1}}} \\
 + {s^2}{\rho _1}{\rho _2}\sum\limits_{k = 0}^{N - M - 1} {\sum\limits_{j = 0}^{N - 1} {{{\left( {\frac{1}{{1 + s{\rho _1}}}} \right)}^{k + 1}}{{\left( {\frac{1}{{1 + s{\rho _2}}}} \right)}^{j + 1}}} }.
\end{multline}
Substituting (\ref{ZF:c5}) into (\ref{ZF:c4}), ${C_{{\gamma _T^{\sf ZF}}}}$ can be computed as
\begin{multline}\notag
{C_{{\gamma _T^{\sf ZF}}}} = \\\frac{{\rho _1}}{{2\ln 2}}{\sum\limits_{k = 0}^{N - M - 1} }{{\cal{I}}_4}
+ \frac{{\rho _2}}{{2\ln 2}}\sum\limits_{j = 0}^{N - 1} {{\cal{I}}_5}
-\frac{{\rho _1}{\rho _2}}{{2\ln 2}}{ \sum\limits_{k = 0}^{N - M - 1} {\sum\limits_{j = 0}^{N - 1} {{\cal{I}}_6} } }.
\end{multline}
where ${{\cal{I}}_4}= {\int_0^\infty  {\frac{{{e^{ - s}}}}{{{{\left( {1 + s{\rho _1}} \right)}^{k + 1}}}}ds} }$, ${{\cal{I}}_5}= {\int_0^\infty  {\frac{{{e^{ - s}}}}{{{{\left( {1 + s{\rho _2}} \right)}^{j + 1}}}}ds} }$ and ${{\cal{I}}_6} = {\int_0^\infty  {\frac{{{e^{ - s}}s}}{{{{\left( {1 + s{\rho _1}} \right)}^{k + 1}}{{\left( {1 + s{\rho _2}} \right)}^{j + 1}}}}ds} } $.
Then, utilizing \cite[Eq. (9.211.4)]{Tables}, we have ${{\cal{I}}_4} = \frac{1}{{{\rho _1}}}\Psi \left( {1,1 - k;\frac{1}{{{\rho _1}}}} \right)$ and ${{\cal{I}}_5} = \frac{1}{{{\rho _2}}}\Psi \left( {1,1 - j;\frac{1}{{{\rho _2}}}} \right)$.
As for ${\cal{I}}_6$, with the help of the identity ${\left( {1 + \beta x} \right)^{ - \alpha }} = \frac{1}{{\Gamma \left( \alpha  \right)}}{\mathop{\rm G}\nolimits} _{1,1}^{1,1}\left( \beta x\middle| \substack{{1 - \alpha }\\
0} \right)$ and \cite[Eq. (2.6.2)]{H-function}, it can be computed as
\begin{align}\notag
{{\cal{I}}_6} &= \int_0^\infty  {s{e^{ - s}}{\mathop{\rm G}\nolimits} _{1,1}^{1,1}\left( {\rho _1}s\middle| \substack{
{ - k}\\
0} \right)} {\mathop{\rm G}\nolimits} _{1,1}^{1,1}\left( {\rho _2}s\middle| \substack{
{ - j}\\
0} \right)ds \notag\\
&= {\mathop{\rm G}\nolimits} _{1,[1:1],0,[1:1]}^{1,1,1,1,1}\left(^{{\rho _1}}_{{\rho _2}}
\middle| \substack{
2\\
{ - k; - j}\\
 - \\
{0;0}} \right),
\end{align}

Finally, pulling everything together, ${C_{{\gamma _T^{\sf ZF}}}}$ is given by
\begin{multline}\label{ZF:c11}
{C_{{\gamma _T^{\sf ZF}}}}= \left( {\sum\limits_{k = 0}^{N - M - 1} {\Psi ( 1,1 - k;\frac{1}{{{\rho _1}}})} } + \sum\limits_{j = 0}^{N - 1} {\Psi (1,1 - j;\frac{1}{{{\rho _2}}})} \right.\\
\left. { - {\rho _1}{\rho _2}\sum\limits_{k = 0}^{N - M - 1} {\sum\limits_{j = 0}^{N - 1} {{\mathop{\rm G}\nolimits} _{1,[1:1],0,[1:1]}^{1,1,1,1,1}\left(^{{\rho _1}}_{{\rho _2}}
\middle| \substack{
2\\
{ - k; - j}\\
 - \\
{0;0}} \right)} } } \right)\frac{1}{{2\ln 2}}.
\end{multline}

Now, substituting (\ref{ZF:c10}) and (\ref{ZF:c11}) into (\ref{ZF:c1}), we obtain the desired result.

\section{Proof for the MMSE/MRT Scheme}
\subsection{Proof of Theorem \ref{theorem:4}}\label{appendix:theorem:4}
Combining (\ref{MMSE:3}) and (\ref{MMSE:4}), the ergodic capacity upper bound of the MMSE/MRT scheme can be computed by
\begin{multline}
C_{\sf MMSE}^{\sf up} = C_{\gamma _1^{\sf MMSE}}+C_{\gamma _2^{\sf MMSE}}-\\
\frac{1}{2}{\log _2}\left( {1 + {e^{{\mathop{\rm E}\nolimits} \left( {\ln {\gamma _1^{\sf MMSE}}} \right)}} + {e^{{\mathop{\rm E}\nolimits} \left( {\ln {\gamma _2^{\sf MMSE}}} \right)}}} \right).
\end{multline}
Noticing that ${\gamma _2^{\sf MMSE}} = {\gamma _2^{\sf MRC}}$, we have ${C_{\gamma _2^{\sf MMSE}} = C_{{\gamma _2^{\sf MRC}}}}$ and ${{\mathop{\rm E}\nolimits} \left( {\ln {\gamma _2^{\sf MMSE}}} \right)} = {\mathop{\rm E}\nolimits} \left( {\ln {\gamma _2^{\sf MRC}}} \right)$.
Hence, the remaining task is to calculate $C_{\gamma _1^{\sf MMSE}}$ and ${{\mathop{\rm E}\nolimits} \left( {\ln {\gamma _1^{\sf MMSE}}} \right)} $.

\subsubsection{Calculation of  ${C_{{\gamma _1^{\sf MMSE}}}}$}
With the help of the following c.d.f. of ${C_{{\gamma _1^{\sf MMSE}}}}$ presented in \cite{ZHU_TWC}
\begin{multline}\notag
{F _{{\gamma _1^{\sf MMSE}}}}\left( x \right) = 1 - \frac{{\Gamma \left( {N,\frac{x}{{{\rho _1}}}} \right)}}{{\Gamma \left( N \right)}} + \Gamma \left( {M + 1} \right){e^{ - \frac{x}{{{\rho _1}}}}}{\left( {\frac{x}{{{\rho _1}}}} \right)^N} \times \\
\sum\limits_{m = {m_1}}^N {\frac{{\rho _I^{N - m + 1}}{}_2{F_1}(M + 1,N - m + 1;N - m + 2; - \frac{{{\rho _I}}}{{{\rho _1}}}x)}{{\Gamma \left( m \right)\Gamma \left( {N - m + 2} \right)\Gamma \left( {m - N + M} \right)}}},
\end{multline}
and using the same methods as in (\ref{MRC:a1}), ${C_{{\gamma _1^{\sf MMSE}}}}$ can be computed as
\begin{multline}\label{MMSE:a1}
{C_{{\gamma _1^{\sf MMSE}}}} = \frac{{{\cal I}_7}}{{2\ln 2}}- \frac{1}{{2\ln 2}}\frac{{\Gamma \left( {M + 1} \right)}}{{\rho _1^N}}\\ \times \sum\limits_{m = {m_1}}^N {\frac{{\rho _I^{N - m + 1}}{{\cal I}_8}}{{\Gamma \left( m \right)\Gamma \left( {N - m + 2} \right)\Gamma \left( {m - N + M} \right)}}},
\end{multline}
where ${{\cal I}_7} = {\int_0^\infty  {\frac{1}{{\Gamma \left( N \right)}}\left( {1 + x} \right)^{-1}{\Gamma \left( {N,\frac{x}{{{\rho _1}}}} \right)}dx} }$, ${{\cal I}_8} = {\int_0^\infty  {\frac{{e^{ - \frac{x}{{{\rho _1}}}}}{x^N}}{{\left( {1 + x} \right)}}{}_2{F_1}( M + 1,N - m + 1;N - m + 2; - \frac{{{\rho _I}}}{{{\rho _1}}}x )dx} } $.

With the help of \cite[Eq. (8.352.4)]{Tables} and \cite[Eq. (3.383.10)]{Tables}, ${\cal I}_7$ can be expressed as
\begin{align}\label{MMSE:a2}
{{\cal I}_7} = {e^{\frac{1}{{{\rho _1}}}}}\sum\limits_{k = 0}^{N - 1} {{{\left( {\frac{1}{{{\rho _1}}}} \right)}^k}\Gamma \left( { - k,\frac{1}{{{\rho _1}}}} \right)}.
\end{align}

Now, let us focus on the computation of ${\cal I}_8$. We first note that, according to \cite[Eq. (9.34.7)]{Tables}, the following equation holds,
\begin{multline}\label{MMSE:a4}
{}_2{F_1}\left( {M + 1,N - m + 1;N - m + 2; - \frac{{{\rho _I}}}{{{\rho _1}}}x} \right) = \\
\frac{{\Gamma \left( {N - m + 2} \right)}{\rho _I}x {\mathop{\rm G}\nolimits} _{2,2}^{1,2}\left( \frac{{{\rho _I}}}{{{\rho _1}}}x\middle| \substack{
{ - M - 1,m - N - 1}\\
{ - 1,m - N - 2}} \right)}{{\Gamma \left( {M + 1} \right)\Gamma \left( {N - m + 1} \right)}\rho _1}.
\end{multline}
Hence, ${\cal I}_8$ can be alternatively expressed as
\begin{multline}\notag
{{\cal I}_8} = \frac{{\Gamma \left( {N - m + 2} \right)}}{{\Gamma \left( {M + 1} \right)\Gamma \left( {N - m + 1} \right)}}\frac{{{\rho _I}}}{{{\rho _1}}}\times\\
\int_0^\infty  {{e^{ - \frac{x}{{{\rho _1}}}}}{x^{N + 1}}{\mathop{\rm G}\nolimits} _{1,1}^{1,1}\left( x\middle| \substack{
0\\
0} \right){\mathop{\rm G}\nolimits} _{2,2}^{1,2}\left( \frac{{{\rho _I}}}{{{\rho _1}}}x\middle| \substack{
{ - M - 1,m - N - 1}\\
{ - 1,m - N - 2}} \right)dx}.
\end{multline}

With the help of \cite[Eq. (2.6.2)]{H-function}, ${\cal I}_8$ can be finally expressed in compact-form as
\begin{multline}\label{MMSE:a3}
{{\cal I}_8} = \frac{{\Gamma \left( {N - m + 2} \right)}}{{\Gamma \left( {M + 1} \right)\Gamma \left( {N - m + 1} \right)}}{\rho _I}\rho _1^{N + 1}\\
\times{\mathop{\rm G}\nolimits} _{1,[1:2],0,[1:2]}^{1,1,2,1,1}\left(^{{\rho _1}}_{{\rho _I}}\middle| \substack{
{N + 2}\\
{0;\left( { - M - 1,m - N - 1} \right)}\\
 - \\
{0;\left( { - 1,m - N - 2} \right)}} \right).
\end{multline}

To this end, substituting (\ref{MMSE:a2}) and (\ref{MMSE:a3}) into (\ref{MMSE:a1}), we have
\begin{multline}\label{MMSE:a17}
{C_{{\gamma _1^{\sf MMSE}}}}
= \frac{e^{\frac{1}{{{\rho _1}}}}}{{2\ln 2}}\sum\limits_{k = 0}^{N - 1} {{{\left( {\frac{1}{{{\rho _1}}}} \right)}^k}\Gamma \left( { - k,\frac{1}{{{\rho _1}}}} \right)}- \frac{{{\rho _1}}}{{2\ln 2}}\times\\
\sum\limits_{m = {m_1}}^N {\frac{{\rho _I^{N - m + 2}} {\mathop{\rm G}\nolimits} _{1,[1:2],0,[1:2]}^{1,1,2,1,1}\left(^{{\rho _1}}_{{\rho _I}}\middle| \substack{
{N + 2}\\
{0;\left( { - M - 1,m - N - 1} \right)}\\
 - \\
{0;\left( { - 1,m - N - 2} \right)}} \right)}{{\Gamma \left( m \right)\Gamma \left( {N - m + 1} \right)\Gamma \left( {m - N + M} \right)}}}.
\end{multline}

\subsubsection{Calculation of  ${{\mathop{\rm E}\nolimits} \left( {\ln {\gamma _1^{\sf MMSE}}} \right)} $}
Similar to the MRC/MRT scheme, we first work out the general moment of ${\gamma _1^{\sf MMSE}}$. According to (\ref{MRC:a8}), we have (\ref{MMSE:a5}) shown on the top of the next page.
\begin{figure*}
\begin{align}\label{MMSE:a5}
&{\rm{E}}\left( {{{\left( {\gamma _1^{\sf MMSE}} \right)}^n}} \right) = n\underbrace {\int_0^\infty  {{x^{n - 1}}\frac{{\Gamma \left( {N,\frac{x}{{{\rho _1}}}} \right)}}{{\Gamma \left( N \right)}}dx} }_{{{\cal I}_9}} - \frac{{\Gamma \left( {M + 1} \right)}}{{\rho _1^N}}\sum\limits_{m = {m_1}}^N {\frac{{\rho _I^{N - m + 1}}}{{\Gamma \left( m \right)\Gamma \left( {N - m + 2} \right)\Gamma \left( {m - N + M} \right)}}} \notag\\
 &\times n\underbrace {\int_0^\infty  {{e^{ - \frac{x}{{{\rho _1}}}}}{x^{N + n - 1}}{}_2{F_1}\left( {M + 1,N - m + 1;N - m + 2; - \frac{{{\rho _I}}}{{{\rho _1}}}x} \right)dx} }_{{{\cal I}_{10}}}.
\end{align}
\hrule
\end{figure*}

Using \cite[Eq. (8.352.4)]{Tables}, ${\cal I}_9$ can be alternatively given by
\begin{align}\label{MMSE:a6}
{{\cal I}_9} = \sum\limits_{k = 0}^{N - 1}{\frac{\int_0^\infty  {{e^{\frac{x}{{{\rho _1}}}}}{x^{k + n - 1}}dx}}{{{k!\rho _1^k}}}}  = \rho _1^n\sum\limits_{m = 0}^{N - 1} {\frac{{\Gamma \left( {m + n} \right)}}{{\Gamma \left( {m + 1} \right)}}}.
\end{align}
Then, using (\ref{MMSE:a4}) and
\cite[Eq. (7.813.1)]{Tables}, ${\cal I}_{10}$ can be finally expressed in compact-form as
\begin{multline}\label{MMSE:a7}
{{\cal I}_{10}} = \frac{{\Gamma \left( {N - m + 2} \right)}}{{\Gamma \left( {M + 1} \right)\Gamma \left( {N - m + 1} \right)}}{\rho _I}\rho _1^{N + n}\\
\times{\mathop{\rm G}\nolimits} _{3,2}^{1,3}\left( {\rho _I}\middle| \substack{{ - N - n, - M - 1,m - N - 1}\\
{ - 1,m - N - 2}} \right).
\end{multline}

To this end, substituting (\ref{MMSE:a6}) and (\ref{MMSE:a7}) into (\ref{MMSE:a5}), we obtain the general moment of ${\gamma _1^{\sf MMSE}}$ as
\begin{multline}\label{MMSE:a14}
{\rm{E}}\left( {{{\left( {\gamma _1^{\sf MMSE}} \right)}^n}} \right) = \underbrace {n\rho _1^n\sum\limits_{m = 0}^{N - 1} {\frac{{\Gamma \left( {m + n} \right)}}{{\Gamma \left( {m + 1} \right)}}} }_{{s_3}\left( n \right)} \\
- \underbrace {\sum\limits_{m = {m_1}}^N {\frac{{\rho _I^{N - m + 2}}n{T_2}(n)}{{\Gamma \left( m \right)\Gamma \left( {N - m + 1} \right)\Gamma \left( {m - N + M} \right)}}}}_{{s_4}\left( n \right)},
\end{multline}
where ${T_2}(n) = \rho _1^n{\mathop{\rm G}\nolimits} _{3,2}^{1,3}\left( {\rho _I}\middle| \substack{{ - N - n, - M - 1,m - N - 1}\\
{ - 1,m - N - 2}} \right)$.

Then, according to (\ref{MRC:a7}), the expectation of ${\ln {\gamma _1^{\sf MMSE}}}$ can be computed as
\begin{align}\label{MMSE:a9}
{\mathop{\rm E}\nolimits} \left( {\ln {\gamma _1^{\sf MMSE}}} \right) &= {\left. {\frac{{d{\mathop{\rm E}\nolimits} \left( {{{\left( {\gamma _1^{\sf MMSE}} \right)}^n}} \right)}}{{dn}}} \right|_{n = 0}} \notag\\&= {\left. {\frac{{d{s_3}\left( n \right)}}{{dn}}} \right|_{n = 0}} - {\left. {\frac{{d{s_4}\left( n \right)}}{{dn}}} \right|_{n = 0}}.
\end{align}

To compute ${\left. {\frac{{d{s_3}\left( n \right)}}{{dn}}} \right|_{n = 0}}$, we first express ${s_3}\left( n \right)$ as
\begin{align}\label{MMSE:a8}
{s_3}\left( n \right) = \Gamma \left( {n + 1} \right)\rho _1^n + n\rho _1^n\sum\limits_{m = 1}^{N - 1} {\frac{{\Gamma \left( {m + n} \right)}}{{\Gamma \left( {m + 1} \right)}}}.
\end{align}
Taking the derivative of (\ref{MMSE:a8}), and let $n=0$, we have
%
\begin{align}\label{MMSE:a10}
{\left. {\frac{{d{s_3}\left( n \right)}}{{dn}}} \right|_{n = 0}} = \psi \left( N \right) + \ln {\rho _1}.
\end{align}

Next, we focus on the computation of ${\left. {\frac{{d{s_4}\left( n \right)}}{{dn}}} \right|_{n = 0}}$.
Again, it can be shown that ${\left. {\frac{{dn{T_2}\left( n \right)}}{{dn}}} \right|_{n = 0}} = {\left. {{T_2}\left( n \right)} \right|_{n = 0}}$.
Therefore, we get
\begin{multline}\label{MMSE:a11}
{\left. {\frac{{d{s_4}( n )}}{{dn}}} \right|_{n = 0}} = \\ \sum\limits_{m = {m_1}}^N {\frac{{\rho _I^{N - m + 2}}{\mathop{\rm G}\nolimits} _{3,2}^{1,3}\left( {\rho _I}\middle| \substack{
{ - N, - M - 1,m - N - 1}\\
{ - 1,m - N - 2}} \right)}{{\Gamma \left( m \right)\Gamma \left( {N - m + 1} \right)\Gamma \left( {m - N + M} \right)}}}.
\end{multline}

To this end, substituting (\ref{MMSE:a10}) and (\ref{MMSE:a11}) into (\ref{MMSE:a9}), we have
\begin{multline}
{\mathop{\rm E}\nolimits} \left( {\ln {\gamma _1^{\sf MMSE}}} \right) = \psi \left( N \right) + \ln {\rho _1} \\
- \sum\limits_{m = m_1}^N {\frac{{\rho _I^{N - m + 2}}{\mathop{\rm G}\nolimits} _{3,2}^{1,3}\left( {\rho _I}\middle| \substack{
{ - N, - M - 1,m - N - 1}\\
{ - 1,m - N - 2}} \right)}{{\Gamma \left( m \right)\Gamma \left( {N - m + 1} \right)\Gamma \left( {m - N + M} \right)}}}.
\end{multline}

Finally, pulling everything together yields the desired result.

\subsection{Proof of Theorem \ref{theorem:5}}\label{appendix:theorem:5}
Combining (\ref{MMSE:3}) and (\ref{MMSE:4}), the ergodic capacity lower bound of the MMSE/MRT scheme can be computed as
\begin{multline}\label{MMSE:a15}
{C_{\sf MMSE}^{\sf low}} = {C_{{\gamma _1^{\sf MMSE}}}} + {C_{{\gamma _2^{\sf MMSE}}}} \\
- \frac{1}{2}{\log _2}\left( {1 + {\mathop{\rm E}\nolimits} \left( {{\gamma _1^{\sf MMSE}}} \right) + {\mathop{\rm E}\nolimits} \left( {{\gamma _2^{\sf MMSE}}} \right)} \right).
\end{multline}
Since ${\gamma _2^{\sf MMSE}} = {\gamma _2^{\sf MRC}}$, we have ${\mathop{\rm E}\nolimits} ( {\gamma _2^{\sf MMSE}}) = {\mathop{\rm E}\nolimits} ( {\gamma _2^{\sf MRC}}) = N{\rho _2}$. Thus, the only thing remains is to compute ${\mathop{\rm E}\nolimits} \left( {{\gamma _1^{\sf MMSE}}} \right)$.

According to the general moment function of ${\gamma _1^{\sf MMSE}}$ in (\ref{MMSE:a14}), it is easy to have
\begin{multline}\label{MMSE:a16}
{\mathop{\rm E}\nolimits} \left( {{\gamma _1^{\sf MMSE}}} \right) = N{\rho _1} \\- {\rho _1}\sum\limits_{m = m_1}^N {\frac{{\rho _I^{N - m + 2}}{\mathop{\rm G}\nolimits} _{3,2}^{1,3}\left( {\rho _I}\middle| \substack{
{ - N-1, - M - 1,m - N - 1}\\
{ - 1,m - N - 2}} \right)}{{\Gamma \left( m \right)\Gamma \left( {N - m + 1} \right)\Gamma \left( {m - N + M} \right)}}}.
\end{multline}

To this end, we can obtain the desired result by pulling everything together.

\nocite{*}
\bibliographystyle{IEEE}

\begin{thebibliography}{1}

\bibitem{AKYILDIZ}
I. Akyildiz, D. M. Gutierrez-Estevez and E. C. Reyes, ``The evolution of 4G systems: LTE-Advanced,'' {\em Physical Communication}, vol. 3, no. 4, pp. 217--244, 2010.

\bibitem{GUO}
W. Guo and T. O'Farrell, ``Relay deployment in cellular networks: Planning and optimization,'' {\em IEEE J. Select. Areas Commun.}, vol. 30, no. 8, pp. 1597--1606, Sept. 2012.

\bibitem{CHO}
J. Cho and Z. J. Haas, ``On the throughput enhancement of down-stream channel in cellular radio networks through multihop relaying,'' {\em IEEE J. Select. Areas Commun.}, vol. 22, no. 7, pp. 1206--1219, Sept. 2004.


\bibitem{TSIFTSIS}
T. A. Tsiftsis, G. K. Karagiannidis, S. A. Kotsopoulos, and F.-N. Pavlidou, ``BER analysis of collaborative dual-hop wireless transmissions,'' {\em Electron. Lett.}, vol. 40, no. 11, pp. 679--681, May 2004.

\bibitem{HASNA}
M. O. Hasna and M.-S. Alouini, ``End-to-end performance of transmission systems with relays over Rayleigh-fading channels,'' {\em IEEE Trans. Wireless Commun.}, vol. 2, no. 6, pp. 1126--1131, Nov. 2003.


\bibitem{FARHADI}
G. Farhadi and N. Beaulieu, ``On the ergodic capacity of wireless relaying systems over Rayleigh fading channels,'' {\em IEEE Trans. Wireless Commun.}, vol. 7, no. 11, pp. 4462--4467, Nov. 2008.

\bibitem{COSTA}
D. B. da Costa and S. A\"{i}ssa, ``Capacity analysis of cooperative system with relay selection in Nakagami-$m$ fading,'' {\em IEEE Commun. Lett.}, vol. 13, no. 9, pp. 637--639, Sept. 2009.

\bibitem{Zhong_TVT}
C. Zhong, M. Matthaiou, G. K. Karagiannidis, A. Huang and Z. Zhang, ``Capacity bounds for AF dual-hop relaying in $\mathcal{G}$ fading channels,'' {\em IEEE Trans. Veh. Technol.}, vol. 61, no. 4, pp. 1730--1740, May 2012.

\bibitem{FAN}
L. Fan, X. Lei and W. Li, ``Exact closed-form expression for ergodic capacity of amplify-and-forward relaying in channel-noise-assisted cooperative networks with relay selection,'' {\em IEEE Commun. Lett.}, vol. 15, no. 3, pp.332--333, Mar. 2011.

\bibitem{SHI}
S. Jin, M. R. McKay, C. Zhong and K.-K. Wong, ``Ergodic capacity analysis of amplify-and-forward MIMO dual-hop systems,'' {\em IEEE Trans. Inf. Theory}, vol. 56, no. 5, pp. 2204--2224, May 2010.

\bibitem{FIRAG}
A. Firag, P. J. Smith, and M. R. McKay, ``Capacity analysis for MIMO two-hop amplify-and-forward relaying systems with the source to destination link,'' {\em IEEE Intl. Conf. Commun. (ICC 2009)}, Dresden, Germany, June 2009, pp. 1--6.

\bibitem{ZHONG_TCOM1}
C. Zhong, S. Jin and K. K. Wong, ``Dual-hop system with noisy relay and interference-limited destination,'' {\em IEEE Trans. Commun.}, vol. 58, no. 6, pp. 764--768, Mar. 2010.

\bibitem{FAWAZ}
F. Al-Qahtani, T. Q. Duong, C. Zhong, K. Qaraqe and H. Alnuweiri, ``Performance analysis of dual-hop AF systems with interference in Nakagami-$m$ fading channels,'' {\em IEEE Sig. Proc. Lett.}, vol. 18, no. 8, pp. 454--457, Aug. 2011.

\bibitem{HIMAL_TVT}
H. A. Suraweera, D. S. Michalopoulos, C. Yuen, ``Performance analysis of fixed gain relay systems with a single interferer in Nakagami-$m$ fading channels,'' {\em IEEE Trans. Veh. Technol.}, vol. 61, no. 3, pp. 1457--1463, Mar. 2012.

\bibitem{SALAMA}
S. Ikki and S. A{\"i}ssa, ``Performance analysis of dual-hop relaying systems in the presence of co-channel interference,'' in {\em Proc. IEEE GLOBECOM
2010}, Miami, FL, Dec. 2010, pp. 1--5.

\bibitem{KRIKIDIS}
I. Krikidis, J. Thompson, S. McLaughlin, and N. Goertz, ``Max-min relay selection for legacy amplify-and-forward systems with interference,'' {\em IEEE Trans. Wireless Commun.}, vol. 8, no. 6, pp. 3016--3027, Jun. 2009.

\bibitem{ZHONG_TCOM2}
C. Zhong, H. A. Suraweera, A. Huang, Z. Zhang, and C. Yuen, ``Outage probability of dual-hop multiple antenna AF relaying systems with interference,'' {\em IEEE Trans. Commun.}, vol. 61, no. 1, pp. 108--119, Jan. 2013.

\bibitem{ZHU_TWC}
G. Zhu, C. Zhong, H. A. Suraweera, Z. Zhang and C. Yuen, ``Outage probability of dual-hop multiple antenna AF systems with linear processing in the presence
of co-channel interference,'' {\em IEEE Trans. Wireless Commun.}, Accepted [Online], Available: http://arxiv.org/abs/1401.1011

\bibitem{WAQAR}
O. Waqar, M. Ghogho, and D. McLernon, ``Outage and ergodic capacity expressions for fixed-gain relay networks in the presence of interference,'' in {\em Proc. IEEE Intl. Conf. Commun. Systems (ICCS 2010)}, Singapore, Nov. 2010, pp. 356--360.

\bibitem{TRIGUI1}
I. Trigui, S. Affes, and A. Stephenne, ``Ergodic capacity analysis for interference-limited AF multi-hop relaying channels in Nakagami-$m$ fading,'' {\em IEEE Trans. Commun.}, vol. 61, no. 7, pp 2726--2734, July 2013.

\bibitem{TRIGUI2}
I. Trigui, S. Affes, and A. Stephenne, ``On the ergodic capacity of amplify-and-forward relay channels with interference in Nakagami-$m$ fading,'' {\em IEEE Trans. Commun.}, vol. 61, no. 8, pp. 3136--3145, Aug. 2013.

\bibitem{HUANG}
Y. Huang, C. Li, C. Zhong, J. Wang, Y. Cheng and Q. Wu, ``On the capacity of dual-hop multiple antenna AF relaying systems with feedback delay and CCI,'' {\em IEEE Commun. Lett.}, vol. 17, no. 6, pp. 1200--1203, June 2013.

\bibitem{ZHU_CONF}
G. Zhu, C. Zhong, H. A. Suraweera, Z. Zhang and C. Yuen, ``Ergodic capacity analysis of dual-hop ZF/MRT relaying systems with co-channel interference,'' in {\em Proc. Intl. Conf. Wireless Commun. and Signal Proces. (WCSP 2013)}, Hangzhou, China, Oct. 2013, pp. 1--6,.

\bibitem{W.Zhang}
W. Zhang, X. Ma, B. Gestner, and D. V. Anderson, ``Designing low-complexity equalizers for wireless systems,'' {\em IEEE Commun. Mag.}, vol. 47, pp. 56--64, Jan. 2009.


\bibitem{H.Ding}	
H. Ding, C. He, and L. Jiang, ``Performance analysis of fixed gain MIMO relay systems in the presence of co-channel interference,'' {\em IEEE Commun. Lett.}, vol. 16, no. 7, pp. 1133--1136, July. 2012.


%
%
%


%

\bibitem{Tables} I. S. Gradshteyn and I. M. Ryzhik, {\em Tables of intergrals,serious and products,} Sixth Edition. San Diago: Acadamic Press, 2000.

\bibitem{R.P.Agrawal}
R. P. Agrawal, ``On certain transformation formulae and Meijer¡¯s G-function of two variables,'' {\em Indian J. Pure Appl. Math.}, vol. 1, no. 4, pp. 537--551, July. 1969.

\bibitem{I.S.Ansari}
I. S. Ansari, S. Al-Ahmadi, F. Yilmaz, M. S. Alouini and H. Yanikomeroglu, ``A new formula for the BER of binary modulations with dual-branch selection over generalized-$K$ composite fading channels,'' {\em IEEE Trans. Commun.}, vol. 59, no. 10, pp. 2654--2658, Oct. 2011.

\bibitem{H.Gao}
H. Gao, P. J. Smith, and M. V. Clark, ``Theoretical reliability of MMSE linear diversity combining in Releigh-fading additive interference channels,'' {\em IEEE Trans. Commun.}, vol. 46, no. 5, pp. 666--672, May. 1998.

{\bibitem{D.Costacluster}
D. B. da Costa and S. A{\"i}ssa, ``Performance analysis of relay selection techniques with clustered fixed-gain relays,'' {\em IEEE Sig. Proc. Lett.}, vol. 17, no. 2, pp. 201-204, Feb. 2010.}

\bibitem{K.Gulati}
{K. Gulati, B. Evans, J. Andrews, and K. Tinsley, ``Statistics of co-channel interference in a field of poisson and poisson-poisson clustered interferers,'' {\em IEEE Trans. Sig. Proc.}, vol. 58, no. 12, pp. 6207--6222, Dec. 2010.}

\bibitem{D.Katselis}
{D. Katselis, ``On estimating the number of co-channel interferers in MIMO cellular systems,'' {\em IEEE Sig. Proc. Lett.}, vol. 18, no. 6, pp. 379--382, June. 2011.}

\bibitem{Y.Ohwatari}
{Y. Ohwatari, N. Miki, T. Asai, T. Abe, and H. Taoka, ``Performance of advanced receiver employing interference rejection combining to suppress inter-cell interference in LTE-Advanced downlink,'' in {\em Proc. IEEE Vehi. Tech. Conf. (VTC Fall 2011)}, San Francisco, USA, Sept. 2011, pp. 1--7.}

\bibitem{R.Narasimhan}
{R. Narasimhan and S. Cheng, ``Channel estimation and co-channel interference rejection for LTE-Advanced MIMO uplink,'' in {\em Proc. IEEE Wirel. Commun. and Netw. Conf. (WCNC 2012)}, Shanghai, China, Apr. 2012, pp. 1--4.}

\bibitem{M. Chiani}
{M. Chiani, M. Win, and H. Shin, ``MIMO networks: The effects of interference,'' {\em IEEE Trans. Inf. Theory}, vol. 56, no. 1, pp. 336--349, Jan. 2010.}


\bibitem{Himal22}
H. A. Suraweera, P. J. Smith and M. Shafi, ``Capacity limits and performance analysis of cognitive radio with imperfect channel knowledge,'' {\em IEEE Trans. Veh. Technol.}, vol. 59, no. 5, pp. 1811-1822, May 2010.


\bibitem{Z.Ding}
Z. Ding, K. K. Leung, D. L. Goeckel and D. Towsley, ``On the application of cooperative transmission to secrecy communications,'' {\em IEEE J. Select. Areas Commun.}, vol. 30, no. 2, pp. 359--368, Feb. 2012.


\bibitem{K.A.Hamdi}
K. A. Hamdi, ``Capacity of MRC on correlated Rician fading channels,'' {\em IEEE Trans. Commun.}, vol. 56, no. 5, pp. 708--711, May. 2008.

\bibitem{H-function} A. M. Mathai and R. K. Saxena, {\em The H-function with Applications in Statistics and Other Disciplines,} New York: Wiley, 1987.



\end{thebibliography}
\begin{footnotesize}

\end{footnotesize}

\end{document}